\documentclass[acmlarge]{acmart}

\usepackage{lscape}
\usepackage{xcolor}
\AtBeginDocument{%
  \providecommand\BibTeX{{%
    \normalfont B\kern-0.5em{\scshape i\kern-0.25em b}\kern-0.8em\TeX}}}

\setcopyright{cc}
\setcctype{by}
\acmJournal{IMWUT}
\acmYear{2025} \acmVolume{9} \acmNumber{2} \acmArticle{28}
\acmMonth{6} \acmDOI{10.1145/3729487}

\begin{document}

\title[RadioGami: Batteryless, Long-Range Wireless Paper Sensors]{RadioGami: Batteryless, Long-Range Wireless Paper Sensors Using Tunnel Diodes}



\author{Imran Fahad}
\email{ifahad@vols.utk.edu}
\orcid{0009-0006-3734-3535}
\affiliation{%
  \institution{University of Tennessee Knoxville}
  \country{USA}
}

\author{Danny Scott}
\orcid{0000-0002-5042-283X}
\affiliation{%
  \institution{University of Tennessee Knoxville}
  \country{USA}
}

\author{Azizul Zahid}
\orcid{0009-0008-0981-5798}
\affiliation{%
  \institution{University of Tennessee Knoxville}
  \country{USA}
}

\author{Matthew Bringle}
\orcid{0000-0003-1457-8519}
\affiliation{%
  \institution{University of Tennessee Knoxville}
  \country{USA}
}

\author{Srinayana Patil}
\orcid{0009-0004-6595-3563}
\affiliation{%
  \institution{University of Tennessee Knoxville}
  \country{USA}
}

\author{Ella Bevins}
\orcid{0009-0008-7385-8522}
\affiliation{%
  \institution{University of Tennessee Knoxville}
  \country{USA}
}

\author{Carmen Palileo}
\orcid{0009-0009-0448-8000}
\affiliation{%
  \institution{University of Tennessee Knoxville}
  \country{USA}
}

\author{Sai Swaminathan}
\email{sai@utk.edu}
\orcid{0009-0009-5055-2632}
\affiliation{%
  \institution{University of Tennessee Knoxville}
  \country{USA}
}

\renewcommand{\shortauthors}{Fahad et al.}


\begin{abstract}
Paper-based interactive RF devices have opened new possibilities for wireless sensing, yet they are typically constrained by short operational ranges. This paper introduces RadioGami, a method for creating long-range, batteryless RF sensing surfaces on paper using low-cost, DIY materials like copper tape, paper, and off-the-shelf electronics paired with an affordable radio receiver (approx. \$20). We explore the design space enabled by RadioGami, including sensing paper deformations like bending, tearing, and origami patterns (Miura, Kresling) at ranges up to 45.73 meters. RadioGami employs a novel ultra-low power (35$\mu$W) switching circuit with a tunnel diode for wireless functionality. These surfaces can sustainably operate by harvesting energy using tiny photodiodes. We demonstrate applications that monitor object status, track user interactions (rotation, sliding), and detect environmental changes. We characterize performance, sensitivity, range, and power consumption with deployment studies. RadioGami advances sustainable, tangible, and batteryless interfaces for embodied interaction.
\end{abstract}





\begin{teaserfigure}
  \includegraphics[width=\textwidth]{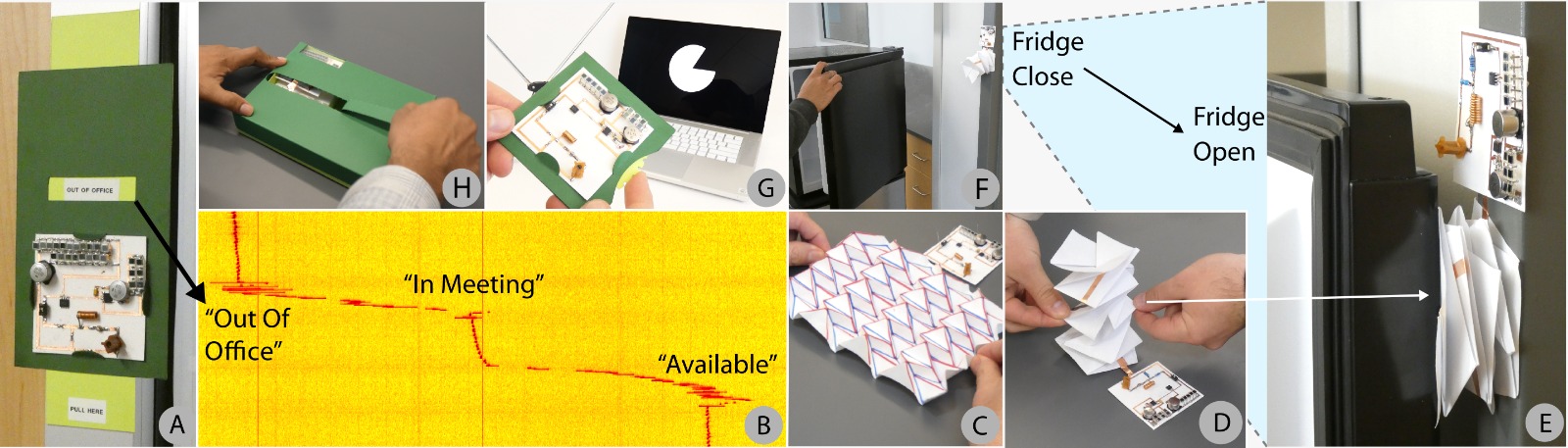}
  \caption{\textbf{Applications and Interactivity of Batteryless, Long-Range Wireless RadioGami Tags.} \textbf{A.} RadioGami tag integrated with a slider mechanism to indicate office presence status (e.g., “Out of Office,” “In Meeting”). \textbf{B.} Monitoring RadioGami tag frequencies using a software-defined radio interface. \textbf{C.} Miura-Ori-inspired surface with a RadioGami tag, demonstrating the combination of origami principles with wireless sensing. \textbf{D.} User interaction with a Kresling origami structure embedded with a RadioGami tag. \textbf{E.} Real-time sensing of refrigerator door activity using a Kresling-based RadioGami tag. \textbf{F.} Kresling origami sensor integrated with a RadioGami tag to detect refrigerator door state during user interaction. \textbf{G.} Tracking rotational motion using a rotary encoder connected to a RadioGami tag. \textbf{H.} Monitoring incremental package tearing with a RadioGami tag.}
  \Description{Applications of RadioGami in interactive and sensing contexts.}
  \label{fig:teaser}
\end{teaserfigure}

\maketitle

\section{Introduction}

Within the HCI research community, there is sustained interest in embedding computational capabilities into everyday materials like paper to create affordable, lightweight, and widely accessible tangible Internet of Things (IoT) devices. Prior research has explored integrating sensing and actuation technologies with paper, enabling deformation sensing \cite{rendl_flexsense_2014,vadgama2017flexy,wessely2018shape} and even actuation \cite{niiyama2015sticky,ogata2015fluxpaper,qi2012animating,wang2018printed}. These approaches illustrate paper's unique properties—flexibility, foldability, and deformability—that make it ideal for creating interactive surfaces. However, one of the primary challenges remains powering these paper-based interfaces.

Using batteries as a power source for widespread deployment introduces notable drawbacks, including ecological costs, ongoing maintenance, and logistical complexities, limiting the vision of ubiquitous computing. Recently, researchers have been developing self-sustainable solutions, moving towards batteryless, wirelessly powered paper-based interfaces. Ambient radio-frequency (RF) signals have been harnessed to power these devices, enabling wireless sensing of user interactions (e.g., touch, speech) without requiring wired connections or batteries \cite{Abowd2020Apr,arora_saturn_2018,li_paperid_2016,gao_livetag_2019}.

These paper-based devices employ radio backscatter communication, offering near-zero-power ($\mu$W) wireless interactivity. They operate by modulating ambient radio waves (e.g., TV bands \cite{liu_ambient_2013}, FM radio \cite{wang_fm_2017}, WiFi channels \cite{Kellogg_wifi_2014}, or RFID tags \cite{smith_rfid-based_2005, philipose_battery-free_2005, li_idsense_2015}) in response to user interactions and reflecting them back to a receiver. However, these devices must often remain close to a powerful emitter (>500 mW) to achieve practical operating ranges, constraining their usability to limited distances (often under 10 m). This proximity requirement restricts the scalability and versatility of wireless, batteryless paper devices for applications such as building-scale interactivity (e.g., spanning multiple floors) or use in non-line-of-sight environments. To address these limitations, we draw on recent work using tunnel diode oscillators (TDOs) to develop RF circuits \cite{varshney_tunnelemitter_2020,eid_58_2020,mir_tunnellifi_2023, reddy_beyond_2023}. TDOs offer a unique capability: they can generate carrier RF signals locally at ultra-low power (just a few $\mu$W), removing the need for external RF emitters and allowing operation over longer ranges.

While recent advances in TDO circuits are promising, they have primarily been demonstrated on rigid PCB substrates \cite{varshney_tunnelscatter_2019,reddy_beyond_2023,varshney_tunnelemitter_2020}, limiting their applicability for the flexible, interactive experiences previously explored with paper in HCI research \cite{li_paperid_2016}. Integrating TDO circuits into paper-based IoT devices to enable long-range, batteryless interactivity remains unexplored. Additionally, while current TDO circuits operate at 56 $\mu$W, there is potential for further power reduction through novel circuit design. Overall, there is a significant knowledge gap in designing and operating long-range, batteryless TDO sensors on flexible, low-cost paper substrates and in advancing circuit designs to minimize power consumption. This is the first work to develop long-range, batteryless wireless sensors on \textit{paper} using TDOs with an ultra-low-power budget of 35uW.

This work introduces RadioGami, a technique for creating novel ultra-low-power TDO circuits on thin, flexible paper substrates. Leveraging the paper’s flexibility, we enable interactive forms, including dynamic folded structures, that alter the TDO circuit’s characteristics through user interactions. These interactions are wirelessly detected via low-cost software-defined radio (SDR) receivers over extended distances. RadioGami combines paper’s unique properties—such as bending, tearing, and complex folding patterns (e.g., Miura, Kresling)—with novel TDO circuitry to introduce a new class of batteryless, long-range "smart paper" interfaces,  as illustrated in Fig.~\ref{fig:teaser}. We detail the fabrication process, covering TDO circuit design, energy harvesting, and power management. Additionally, we present power budget characterizations, operational range data, and results from a real-world deployment study, demonstrating RadioGami’s practical capabilities.

Our key contributions include: 
\begin{itemize}  
\item \textit{RadioGami tags, an end-to-end design of paper-based flexible sensors and interfaces that enable long-range, batteryless interactivity using tunnel diodes}. Our system leverages ultra-low-power TDO circuits to facilitate interaction without the need for batteries, as detailed in \textbf{Section~\ref{sec:System}}. Unlike conventional flexible batteryless wireless sensors, our approach supports operation over long distances (45.7m) while maintaining minimal power consumption (35$\mu$W) on flexible substrates. A comparative analysis of our novel approach is provided in \textbf{Section~\ref{sec:compare}}.

\item \textit{We present novel circuit techniques for controlling TDO circuits in RadioGami tags, including an intermittent power switch utilizing subthreshold MOSFETs.} The novelty of this switch lies in its ability to operate at voltages below 1.0V, significantly reducing power consumption from harvested energy and enabling robust sensing even under tight power budgets (35$\mu$W). The design of this switching technique is detailed in \textbf{Section~\ref{subsec:Switching}}.

\item \textit{We contribute a comprehensive empirical evaluation of our long-range, wireless RadioGami tags powered by ultra-low-power TDO circuits,} as detailed in \textbf{Section~\ref{sec:evaluation}}. The evaluation covers operational range, power consumption, and deformation response, offering key insights into the system’s performance and robustness.

\item \textit{A library of tangible sensor designs that integrate origami and paper-based mechanisms to modulate flexible TDO circuits.} These include Miura-Ori and Kresling folds, paper-based rotary encoders, pull-tab sliders, and more, as illustrated in \textbf{Section~\ref{sec:sensors}}.

\item \textit{We introduce “interaction-activated” RadioGami tags, designed to operate in low-lux environments,} as detailed in \textbf{Section~\ref{sec:low_lux}}. These tags conserve power by remaining dormant until triggered by user interaction. We demonstrate their reliability through a 60+ hour longitudinal deployment in realistic, everyday conditions across varying ambient light levels (30–350 lux).
\end{itemize}

Building on these contributions, our work provides a foundation for developing interactive, batteryless paper interfaces that harness ultra-low-power electronics for sustainable and long-range sensing applications.

\section{Background and Related Work}
Paper substrate is used in flexible electronics due to its lightweight design, cost-effectiveness, mechanical flexibility, and environmental sustainability. Compared to traditional materials like polyimide (PI) \cite{OptoSense_Zhang_2020}, polydimethylsiloxane (PDMS) \cite{physiologicalmeasurement_liu_2021}, silicon \cite{hsieh2019rftouchpads}, metal foils \cite{gao_livetag_2019}, or textiles \cite{Wearable_Lemey_2016}, paper offers significant advantages in terms of recyclability and low production costs. Researchers have embedded several computational technologies such as  sensors, actuators as well as wireless communication into paper substrates and powered them in a batteryless manner, further demonstrating the potential of paper-based electronics.

In this section, we will discuss three key areas of previous works. First, we discuss the RadioGami tag in the context of paper-based interactive sensors and surfaces. Next, we highlight the advantages of power efficiency and communication range of RadioGami tags over existing paper/flexible batteryless wireless sensors. Finally, we examine our tunnel diode oscillator (TDO) design approach compared to state-of-the-art TDOs for batteryless wireless sensing. 

\subsection{Paper Based Interactive Sensors and Surfaces}
Research on making paper interactive has led to various approaches that integrate sensing and actuation capabilities into paper surfaces, enabling a range of tangible IoT interfaces. This work can be categorized based on power requirements and communication capabilities: paper-based sensors without wireless or batteryless capabilities, paper-based sensors with wireless connectivity but reliant on batteries, and finally, batteryless, wireless paper-based sensors that operate at limited ranges.

First, a substantial body of research has focused on enhancing the interactivity of paper through methods like cutting, printing, and drawing to embed sensing elements. Approaches such as Crafting Technology \cite{Crafting_Buechley_2012} and Foldio \cite{foldio_olberding_2015} illustrate the integration of sensors and actuators directly into paper, although these designs are limited to wired power sources. SwellSense \cite{SwellSense_Cheng_2023}, for instance, employs screen-printed stretchable circuits on microcapsule paper to create tactile devices like Braille displays; however, it requires DC power and lacks any wireless capability, restricting its deployment flexibility. Similarly, Print-A-Sketch \cite{Print-A-Sketch_Pourjafarian_2022} utilizes a handheld printer to embed circuits onto surfaces, yet this design remains dependent on direct power sources, which limits its scalability for wireless applications. Electronic Popables \cite{electronicPopables_qi_2010} combines traditional pop-up mechanisms with flexible electronics for interactive books, while Kunihiro et al. \cite{paperwoven_kato_2022} extended this idea by creating 3D structures using conductive paper strips to sense touch and proximity. Despite their interactivity, these systems all rely on wired power, making them unsuitable for untethered or large-scale applications.

In contrast, recent developments have enabled paper-based sensors to incorporate wireless functionality, albeit with the constraint of battery dependency. Origami-inspired designs have also explored wireless functionality. Li et al. developed a Miura-Ori stretchable circuit board for a wireless ECG system, which, however, still relies on battery power \cite{miura_li_2021}. Other designs, like Moeinnia’s \cite{origami_moeinnia_2024} pressure sensor array, require a 5V power bank, while Chen’s \cite{origami_chen_2022} humidity sensor and Karmakar’s \cite{origami_karmakar_2024} folded pressure sensor integrate wireless sensing but remain battery-dependent, impacting sustainability and practicality for extensive deployment.

Efforts to create batteryless, wireless sensors on paper substrates have also advanced the field, though current systems are limited in range. PaperID \cite{li_paperid_2016} exemplifies this category by employing RFID tags to detect touch, material coverage, and movement trajectories without a battery, but its range is restricted to 6-10 meters, and it requires costly RFID readers. Saturn \cite{arora_saturn_2018}, a batteryless paper-based microphone, allows wireless transmission of audio data but has limited range and depends on ambient RF signals. Kim et al. \cite{origami_kim_2022} demonstrated a batteryless wireless pressure sensor using LC circuits with data transmission capabilities up to 10 cm, but this system depends on a Vector Network Analyzer for readouts. Liu’s \cite{origami_liu_2012} self-powered device, while batteryless, requires a digital multimeter for data acquisition, further constraining its practical range and application. Although these advancements mark significant progress in batteryless and wireless paper-based sensors, achieving long-range capabilities remains challenging.

Our work, RadioGami, addresses these limitations by combining the flexibility of paper-based interactive sensors with the long-range(>45 meters), batteryless wireless sensing capabilities provided by tunnel diode oscillators (TDOs). RadioGami is the first system to integrate TDO circuits directly into paper, enabling a variety of sensor types, including sliders, rotary encoders, and origami-inspired surfaces, to achieve long-range wireless interaction. This approach extends the operational range of paper-based interfaces and enables sustainable, batteryless functionality suitable for scalable deployments.

\begin{figure}[!tb]
\includegraphics[width=1\linewidth]{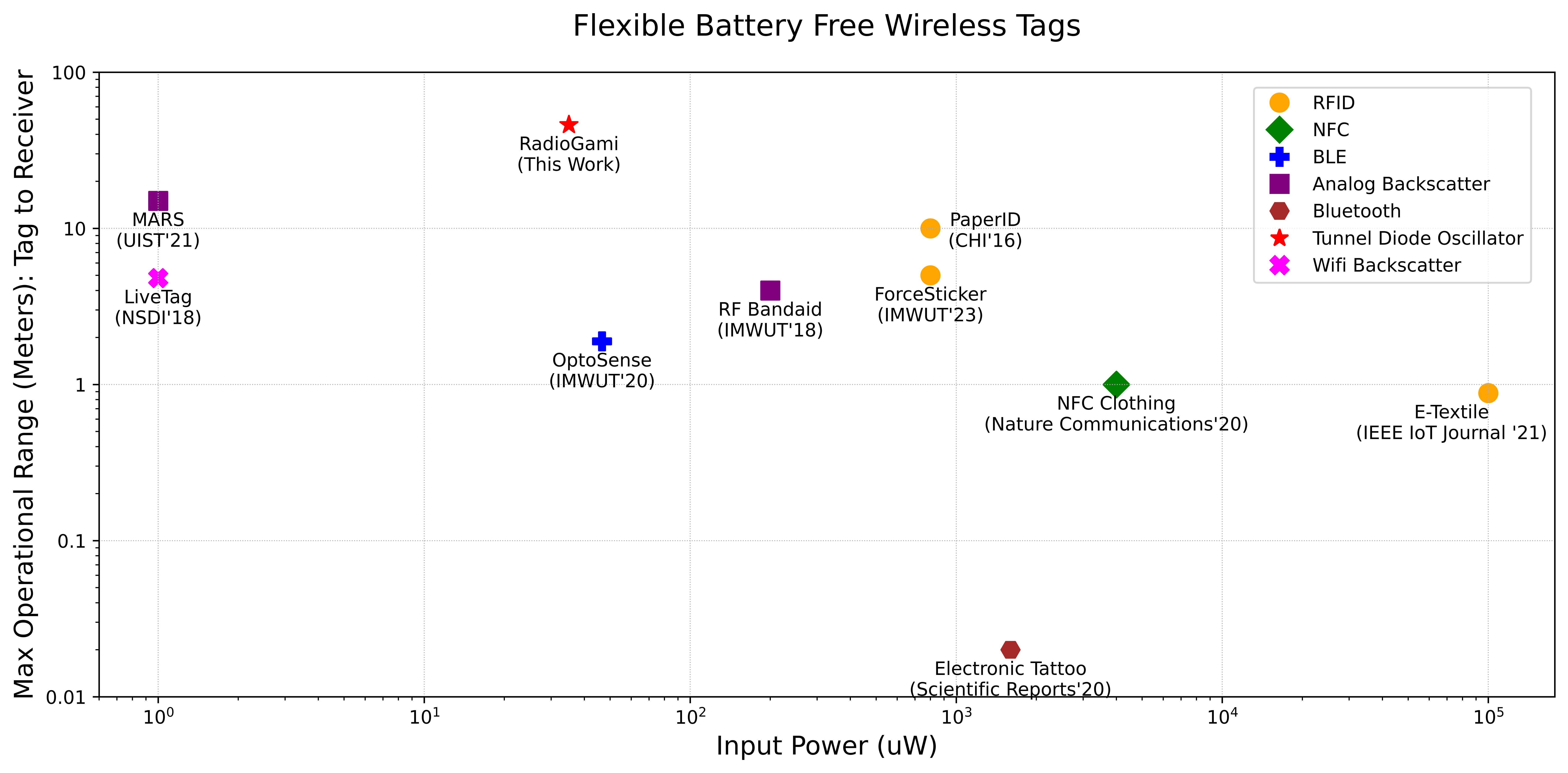}
\centering
\caption{Comparison of power consumption and operational range for flexible batteryless wireless sensing tags such as MARS \cite{arora_mars_2021}, RF Bandaid \cite{ranganathan_rfbandaid_2018}, PaperID \cite{li_paperid_2016}, OptoSense \cite{OptoSense_Zhang_2020}, ForceSticker \cite{ForceSticker_Gupta_2023}, LiveTag \cite{gao_livetag_2019}, Electronic Tattoo \cite{electronictattoo_alberto_2020}, NFC Clothing \cite{NFCclothing_lin_2020}, Sweat cortisol \cite{sweatcortisol_cheng_2021}. The graph plots power consumption (in $\mu$W) against range (in meters) with colored markers representing distinct technologies as indicated in the legend.}
\label{fig:compare}
\centering
\Description[Power vs Range of different Battery-free sensing tags.]{The Fig. provides a comparison among the tags.}
\end{figure}

\subsection{Power vs. Range of Paper/Flexible Batteryless Wireless Sensors} \label{sec:compare}
Batteryless wireless sensing technologies, particularly those using flexible and paper substrates, show distinct trade-offs between input power and operational range. These systems utilize various communication technologies, including RFID, NFC, Bluetooth, analog backscatter, and WiFi backscatter, each optimized for specific applications based on power and range. The following section examines each technology type, comparing input power, operational range, and substrate suitability across analog backscatter, RFID, NFC, Bluetooth, BLE, and WiFi backscatter technologies.

Analog backscatter systems, such as MARS \cite{arora_mars_2021}, operate at a very low power of 1 $\mu W$ with a range of 15 meters and are built on a flexible, low-cost platform comparable to sticky notes. RF Bandaid \cite{ranganathan_rfbandaid_2018}, requiring 200 $\mu W$, supports a range of up to 4 meters and is designed as a disposable, continuous physiological monitoring platform with a bandaid-like substrate.

In RFID-based systems, PaperID \cite{li_paperid_2016} operates at 800 $\mu W$ and reaches up to 10 meters, enabling detection of touch and gesture interactions for applications such as interactive pop-up books and real-time feedback on paper interfaces. Additionally, ForceSticker \cite{ForceSticker_Gupta_2023}, operating at 800 $\mu W$ with a range of 5 meters, is suited for in-vivo force sensing applications in orthopedic implants and packaging integrity checks. The ForceSticker sensor is a flexible, thin polymer layer composed of either Ecoflex or neoprene rubber, allowing it to maintain flexibility in various applications.

NFC and Bluetooth solutions provide close-range communication with low power needs. NFC Clothing \cite{NFCclothing_lin_2020}, operating at 4 mW within a 1-meter range, integrates conductive threads on textile substrates through computer-controlled embroidery, making it well-suited for wearable applications requiring brief, multi-node physiological data sensing, such as monitoring spinal posture, temperature, and gait. Electronic Tattoo \cite{electronictattoo_alberto_2020}, built on an ultrathin (~5 $\mu m$) polymeric film that carries a printed Silver-Indium-Gallium (Ag-In-Ga) conductive circuit, uses Bluetooth and operates at 1.6 mW with a close range of 0.02 meters, for applications such as ECG monitoring.

BLE and WiFi backscatter technologies offer solutions for low-power sensing. OptoSense \cite{OptoSense_Zhang_2020}, which operates at 46.6 $\mu W$ with a 1.89-meter range, utilizes a flexible polyimide substrate combined with off-the-shelf photodetectors, flexible photovoltaic cells, and flexible printed circuits (FPCs) to detect ambient light for user activity and object interaction monitoring. LiveTag \cite{gao_livetag_2019}, using a flexible, paper-like material with a thin metallic conductive layer, uses WiFi backscatter to achieve a 4.8-meter range at just 1 $\mu W$.

This work introduces RadioGami, a flexible, paper-based wireless sensor platform that achieves a range of 45.7 meters with just 35 $\mu W$ of power using 25 tiny photodiodes. RadioGami uses a tunnel diode oscillator (TDO)-based design to provide a low-cost, micro-power solution with long-range and minimal infrastructure requirements. RadioGami also utilizes affordable Software-Defined Radios (SDRs), priced under \$20, to receive non-line-of-sight signals. Fig.~\ref{fig:compare} presents a comparative analysis of the input power and operational range of RadioGami tags within the landscape of paper/flexible batteryless wireless sensors.

\subsection{Tunnel Diode Oscillator and Background}
Recent work on Tunnel Diode Oscillators (TDOs), such as TunnelScatter \cite{varshney_tunnelscatter_2019}, Judo \cite{varshney_judo_2022}, Enabling L3 \cite{yan_enablingL3_2022}, TunnelLiFi \cite{mir_tunnellifi_2023}, and Going Beyond Backscatter \cite{shah_going_2023}, demonstrates batteryless wireless sensors built on conventional rigid printed circuit boards. These sensors exploit the unique electrical properties of tunnel diodes, which exhibit negative differential resistance due to electron tunneling in a heavily doped p-n junction. This property enables efficient, high-frequency operation with low power consumption, making tunnel diodes suitable for oscillators, RF-based sensors, and high-speed switches.

For example, TunnelScatter \cite{varshney_tunnelscatter_2019} harvests ambient energy from carrier signals or light, consuming a peak biasing power of 57 $\mu W$ and achieving a range of 18 meters. It relies on active components, including an MSP430 microcontroller (MCU) for data processing, and requires a receiver with a transimpedance amplifier. Similarly, Judo \cite{varshney_judo_2022} transmits data over distances up to 100 meters with less than 100 $\mu W$ of power, using injection-locking with an external carrier signal and an MCU. TunnelLiFi \cite{mir_tunnellifi_2023} supports visible light-based, line-of-sight communication at micro-power levels, leveraging a TDO.

While RadioGami draws inspiration from prior works, it introduces several critical innovations. Unlike these systems, RadioGami does not rely on an ambient carrier source, MCU, or amplifier for wireless sensing. Instead, it employs a low-voltage intermittent power switch, which is hardware-programmed to control the duty cycle and switching frequency. This design reduces power consumption to 35 $\mu W$ and extends the operational range to over 45 meters. The novel switching circuit utilizes a zero-threshold N-channel MOSFET, with a 555-timer generating the gate pulse. MARS \cite{arora_mars_2021} uses a zero-threshold transistor to implement a low-voltage oscillator, whereas RadioGami leverages the same component as an intermittent switch to enhance power management and optimize operational range. RadioGami is the first to implement a TDO on paper substrates for flexible, batteryless wireless sensing. This innovation extends the operational range of paper-based sensors, supports scalable deployments, and significantly advances the field.

\section{RadioGami System Design and Implementation} \label{sec:System}

This section outlines the design and implementation of the RadioGami system. It begins with an overview of the circuit and system architecture, followed by a detailed explanation of the Tunnel Diode Oscillator (TDO) design and its role in Radio Frequency (RF) transmission. The discussion then addresses the TDO’s power requirements and the considerations for effective biasing. Next, it highlights the ambient power management strategies, including an innovative intermittent power switching mechanism. The section concludes by examining spectrum sensing and FCC compliance, broadcast frequency selection, RF antenna design, receiver setup, and the fabrication process for the RadioGami tag.

\begin{figure}[!tb]
\includegraphics[width=1\linewidth]{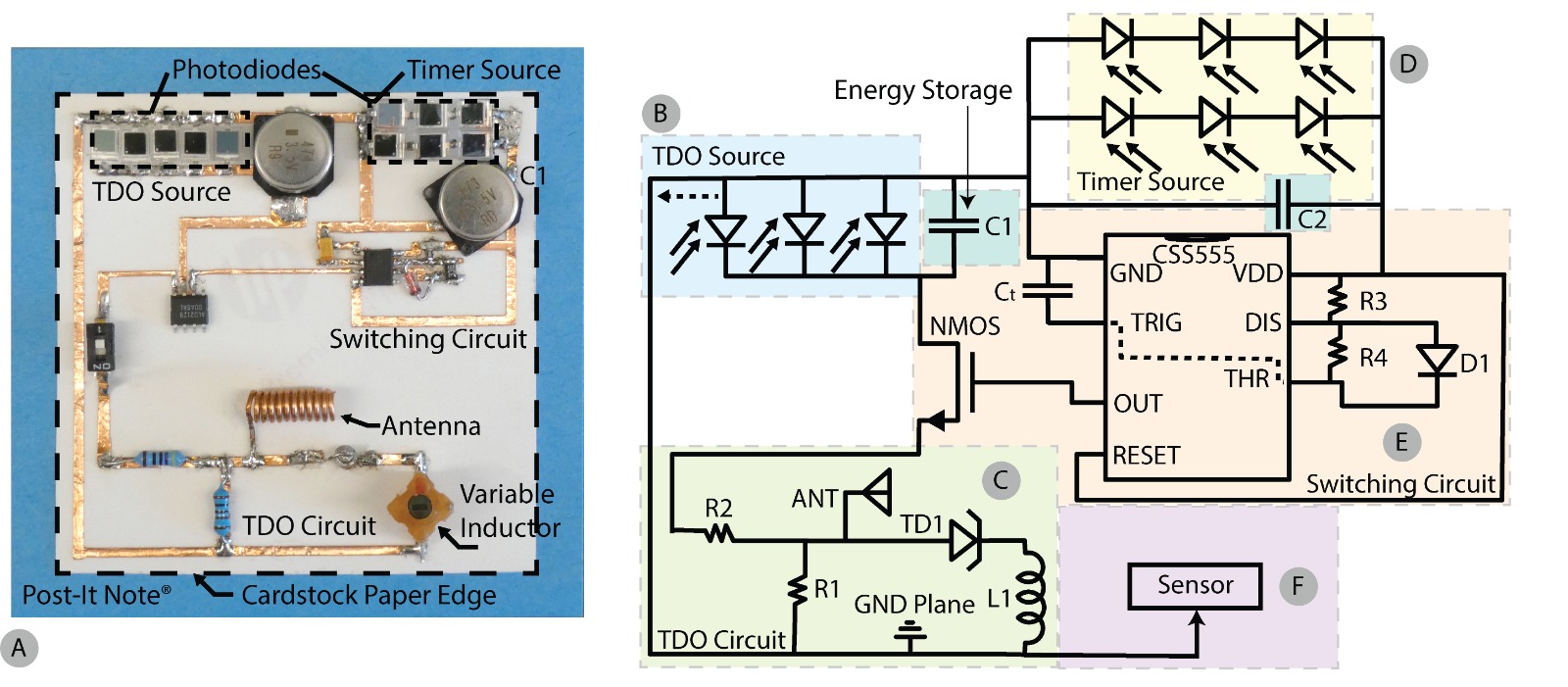}
\centering
\caption{\textbf{RadioGami System Overview.} \textbf{A.} Physical layout of the RadioGami Circuit on Paper. \textbf{B,C,D,E,F.} Circuit components of the RadioGami tag.}
\label{fig:circuit2}
\centering
\Description[Overview of the RadioGami device including physical layout and schematic diagram.]{The figure shows the physical layout of the RadioGami device with areas for optional solar diode expansion and a schematic diagram detailing the paper-based radio device circuit, including energy harvesting components, the tunnel diode oscillator, and connections for RF sensors.}
\end{figure}

\subsection{RadioGami Circuit and System Overview}
The RadioGami tag, depicted in Fig.~\ref{fig:circuit2}, is a batteryless wireless tag created on a paper substrate with copper tape, sized similarly to a Post-It note. As the first implementation of a Tunnel Diode Oscillator (TDO) on a paper substrate, the tag’s circuit integrates several critical sub-components that enable its sustainable operation by harvesting energy from ambient light. To achieve this, we employ a series of small photodiodes (Fig.~\ref{fig:circuit2}B) that capture ambient light, with the harvested energy stored in two supercapacitors, C1 and C2 (Fig.~\ref{fig:circuit2}B, \ref{fig:circuit2}D). These supercapacitors work as energy storage elements for the circuit. The circuit’s core component is a tunnel diode oscillator (Fig.~\ref{fig:circuit2}C), designed to enable RF transmissions using only micropower. A novel switching circuit (Fig.~\ref{fig:circuit2}E) manages power delivery to the oscillator, controlling its duty cycle and frequency based on the energy stored in the supercapacitors C1 and C2. This enables the tag to operate efficiently in varying lighting conditions. Paper sensors and surfaces attached to the RadioGami tag also modulate the TDO’s frequency, allowing for interactive applications. A helical antenna (Fig.~\ref{fig:circuit2}C) enables RF signal broadcasting, which is received via a low-cost software-defined radio (SDR). The received data is utilized for a range of applications. Below, we provide further detail on the working principles of each sub-component in the RadioGami design, explaining their functions and design rationales.

\subsection{Tunnel Diode Oscillator Design}

\begin{figure}[!tb]
\includegraphics[width=\textwidth]{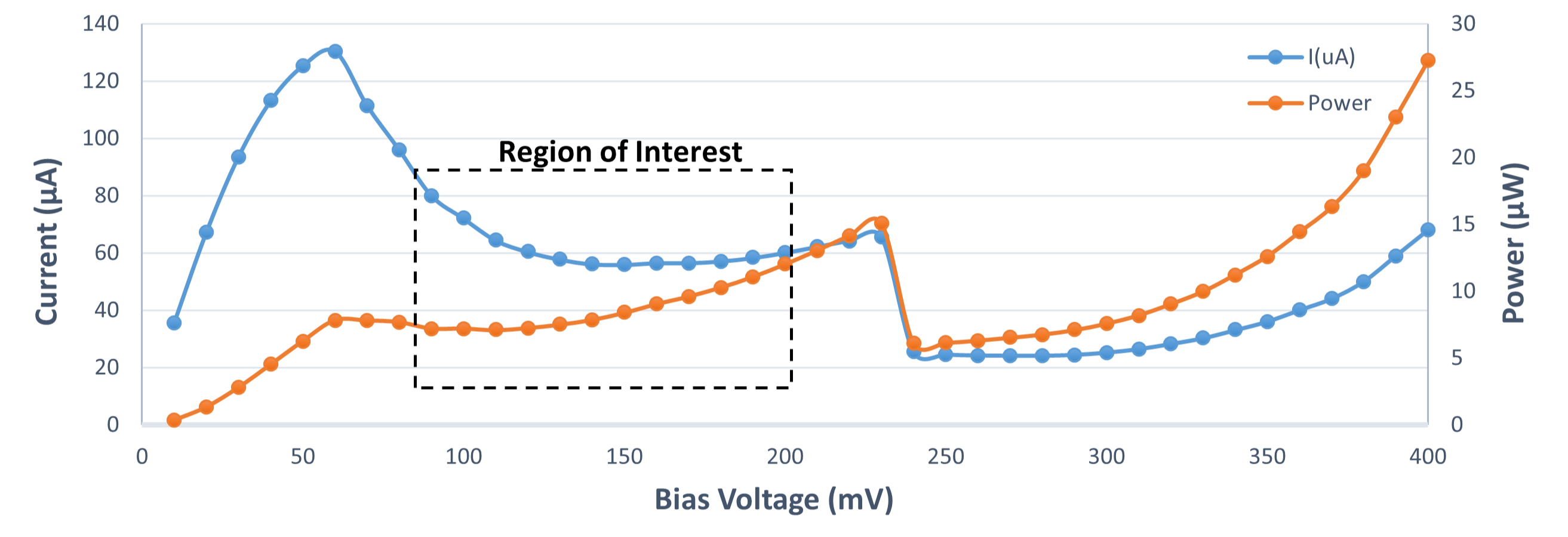}
\centering
\caption{\textbf{Tunnel Diode Characteristics.} \textbf{A.} Measured current-voltage (I-V) characteristics of the MP1X4266 tunnel diode, highlighting its negative differential resistance region. \textbf{B.} Power consumption of the MP1X4266 across varying bias voltages, demonstrating its ultra-low power operation.}

\label{fig:IV}
\Description{power: a Description}
\end{figure}

The Tunnel Diode Oscillator (TDO) enables ultra-low power RF transmission in the RadioGami tag. We implement this design on a flexible paper substrate, where photodiodes harvest ambient light to power the circuit. The oscillator incorporates the MP1X4266 tunnel diode \cite{mpulse_tunnel_diode} and operates within its negative differential resistance region (65–200 mV, Fig.~\ref{fig:IV}) to sustain stable oscillations. To achieve impedance matching and enable high-frequency operation, we design the TDO using a small-signal model that incorporates the tunnel diode’s negative conductance ($g_d$), the equivalent resistance of resistors R1 and R2 ($R_T$), junction capacitance ($C$), and inductance ($L$). To ensure stable oscillations, the tunnel diode is biased within its negative differential resistance (Fig.~\ref{fig:IV}) region using a resistor network consisting of R1 and R2 (Fig.~\ref{fig:circuit2}C). This network establishes the operating point of the diode, ensuring that it remains within the required voltage range for negative differential resistance. We integrate this biasing network with an external inductance (L) to form a resonant circuit that determines the oscillation frequency.

The oscillation frequency and stability depend on the interplay between the tunnel diode’s negative differential conductance and the resonant circuit parameters. The following equations provide approximate relationships among the parameters:

\begin{equation} \label{eq:bias}
\frac{{R_T}}{{\lvert \text{g}_d \rvert}} = \frac{L}{C}
\end{equation}

\begin{equation} \label{eq:resonant_freq}
f_o = \frac{1}{2\pi} \cdot \sqrt{\frac{1 - R_T \cdot \lvert \text{g}_d \rvert}{L \cdot C}}
\end{equation}

These equations \cite{lowry_general_1961} serve as an initial guide for component selection. However, we empirically fine-tune the parameters to optimize oscillation stability. Given the negative differential conductance ($g_d$) of -0.01 Mho ($\mho$) in our chosen tunnel diode, we select R1 = 1 k$\Omega$ and R2 = 470 $\Omega$ to maintain the biasing voltage within 100–200 mV, ensuring stable oscillation. The voltage across R1 stabilizes at 150 mV, placing the diode in its optimal operating region. To adjust the oscillation frequency within the target RF band, we use a tunable Coilcraft Variable Inductor (146 Unicoil Tunable) as the inductive component (see Fig.~\ref{fig:circuit2}C). This allows fine-tuning of the inductance to achieve precise frequency control.

\subsection{Power Harvesting and Energy Storage}
RadioGami achieves batteryless operation by harvesting ambient light energy through photodiodes (BPW34 \cite{bpw}). These photodiodes convert ambient light into electrical energy, stored in two supercapacitors, C1 and C2, each catering to distinct power requirements. The photodiodes enable efficient energy harvesting under typical indoor lighting conditions.

To address the specific requirements of the TDO oscillator and the switching circuit, the system utilizes a dual power source design to ensure reliable and sustained operation. The 0.47 F supercapacitor, C1, serves as a primary source, maintaining a stable 250 mV to bias the TDO within its negative differential resistance region (65 mV to 200 mV), enabling consistent oscillations. This is complemented by an auxiliary source that supports the switching circuit during demand. The design balances charge retention and rapid initial charging, allowing the system to withstand brief light interruptions, such as passing shadows. Under typical indoor lighting conditions, C1 charges to 250 mV within 3 minutes and 26 seconds, ensuring uninterrupted TDO functionality.

Meanwhile, the 0.047 F supercapacitor, C2, charged to 1 V, powers the switching circuit, which demands minimal current (7 $\mu$A) but operates at a higher voltage. This setup employs two strings of three photodiodes, enabling C2 to accumulate charge gradually while maintaining a stable supply for intermittent switching.

\subsection{Power Availability and Management} \label{subsec:Power}

\begin{figure}[tb]
\includegraphics[width=1\linewidth]{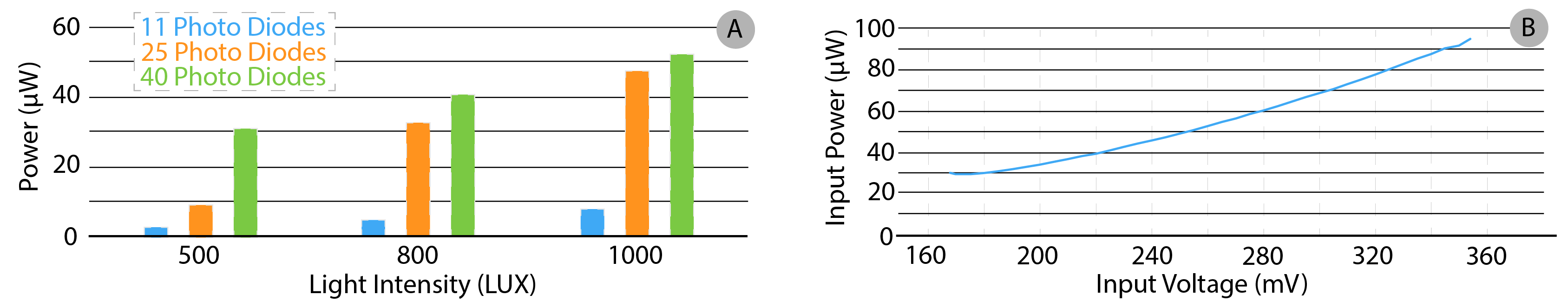}
\centering
\caption{\textbf{TDO Circuit Power Characterization.} \textbf{A.} Power output of photodiodes under varying illumination levels and configurations, demonstrating the impact of increasing light intensity on energy harvesting. \textbf{B.} Input power as a function of voltage, indicating the operational range required for stable circuit performance.}
\label{fig:power}
\centering
\Description[Power Analysis]{Power characterization of the TDO circuit using photodiodes under different lighting conditions and input voltages.}
\end{figure}

The TDO circuit operates with an input power requirement of 50 $\mu$W at a bias voltage of 250 mV (Fig.~\ref{fig:power}B). To evaluate photodiode power generation under controlled illumination, we examined configurations of 11, 25, and 40 photodiodes at 500, 800, and 1000 lux (Fig.~\ref{fig:power}A). These levels were selected based on the Illuminating Engineering Society (IES) and the US National Research \& Development Center guidelines, which recommend a minimum of 500 lux for office spaces, laboratories, and libraries \cite{eye_li_2018, IES_Lighting_Library}. The 800- and 1000-lux conditions represent brighter indoor environments with additional artificial lighting, while the 500-lux level reflects typical office conditions. We did not include a 300-lux setting, as it falls below the IES standard and is primarily used for testing low-light conditions, according to Li et al. \cite{eye_li_2018}, pg. 13.

Fig.~\ref{fig:power}A presents the measured power output across these conditions. At 500 lux, configurations with 11, 25, and 40 photodiodes generated 2.5 $\mu$W, 8 $\mu$W, and 27 $\mu$W, respectively. At 800 lux, power output increased to 10 $\mu$W, 30 $\mu$W, and 40 $\mu$W, while at 1000 lux, it reached 12 $\mu$W, 45 $\mu$W, and 60 $\mu$W. These results confirm that higher lux levels improve power harvesting, but the generated power remains insufficient to meet the 50 $\mu$W requirement at lower illumination levels with fewer photodiodes.  Since the available power at 500 and 800 lux does not meet the system’s operational requirement in all configurations, we implemented an energy-efficient power management strategy, detailed in Section~\ref{subsec:Switching}.

\subsection{Intermittent Power Switching} \label{subsec:Switching}
To bridge the gap between harvested and required power, we implemented an "Intermittent Power Switching" strategy, which reduces the TDO’s average power consumption by periodically toggling power delivery on and off. This circuit, built around an NMOS transistor and a CSS555 timer, facilitates intermittent operation to conserve energy. Powered by the 0.047 F supercapacitor, C1, the circuit employs a low-threshold NMOS transistor with minimal on-resistance (14 ohms) to control switching efficiently. The CSS555 timer generates gate pulses to regulate the MOSFET, enabling precise power delivery adjustments based on photodiode configuration and ambient light levels.

To ensure consistent operation, the 555 timer requires a minimum input voltage of 0.6 V, which is achieved by configuring three photodiodes in series to generate approximately 0.9 V. To further stabilize the voltage against minor light fluctuations, we added two parallel strings of photodiodes, each comprising three photodiodes in series. This arrangement ensures a steady and reliable voltage supply. The energy stored in C1 enables intermittent switching with a customizable switching frequency and duty cycle, significantly reducing the average power consumption to well below the 50 $\mu$W required for continuous operation.

\begin{equation} 
\text{Clock Frequency} : \frac{0.455}{(R3 + 2 \times R4) \times C_T} 
\label{eq:clock_frequency} 
\end{equation}

\begin{equation} 
\text{Duty Cycle}: \frac{R3}{R3 + 2 \times R4} 
\label{eq:duty_cycle} 
\end{equation}

Using equations \ref{eq:clock_frequency} and \ref{eq:duty_cycle}, we first calculate R3, R4 and $C_T$ theoretically. We then set R3 (1 M$\Omega$), R4 (33 M$\Omega$), and $C_T$ (10 $\mu$F) based on practical requirements. We place a bypass diode in parallel with R4 to achieve a 10\% duty cycle at a 60 Hz clock frequency. This configuration minimizes power consumption and allows RadioGami to function effectively even in low-light environments. We provide a comprehensive overview of intermittent switching in Subsection \ref{subsec:Switch}.

\subsection{Spectrum Sensing and FCC Compliance} 
We evaluate compliance with FCC Part 15 \cite{fcc} regulations for TV white space (TVWS) devices by conducting spectrum sensing and transmitter power measurements. The FCC regulates TVWS by categorizing devices into two main types: fixed and lower-power "personal/portable" devices, aiming to protect licensed users. Personal/portable devices are restricted to operating within TV channels 21-36 (512-608 MHz) and 38-51 (614-698 MHz), with a maximum Effective Isotropic Radiated Power (EIRP) of 100 mW, or 40 mW if adjacent to an occupied channel \cite{whitespace_ramjee_2016}. These devices function in two modes: Mode I devices depend on other devices for channel information, while Mode II devices independently utilize geolocation and access TV band databases to select available channels. The FCC mandates periodic location updates and immediate cessation of operation if channels become unavailable, ensuring minimal interference with primary users.

To assess compliance, we performed spectrum sensing using a TinySA spectrum analyzer with wideband antennas, scanning the 500-600 MHz range at 50 kHz intervals over a 24-hour period in a laboratory environment. This scan revealed dense occupancy in the 505-560 MHz band, while the 575-600 MHz band showed minimal activity, indicating a suitable range for operation. We then measured the RadioGami tag's transmitter power at a distance of 0.01 meters using the TinySA spectrum analyzer, recording a transmitted power of -50 dBm (10 µW), well below the FCC's limit of 40 mW. This ensures the RadioGami tag remains within full regulatory compliance. Based on these findings, we selected the 575-600 MHz range within the UHF spectrum for operation.

\subsection{RF Antenna Design and Receiver} \textbf{Helical Antenna Design.} With the 575-600 MHz range defined, we designed a normal-mode helical antenna to meet the tag’s broadcasting requirements. The helix configuration’s circumference is significantly shorter than a full wavelength, and its pitch is less than a quarter wavelength, enabling it to function as a quarter-wavelength monopole antenna \cite{helical}. At the selected frequency, the corresponding wavelength ranges between 52.17 cm and 50 cm. Using polyurethane-enameled copper wire, 22 AWG in diameter and 12 cm in length, we wound the wire into 14 turns to form the helical structure with a diameter of 4.5 mm (see Fig.~\ref{fig:circuit2}A). This results in an omnidirectional radiation pattern, which supports the broadcasting needs of our application.

\textbf{RF Receiver.} To receive the RF signals from our tags, we use an RTL-SDR (Model: V3) software-defined radio (SDR) connected to a dipole antenna \cite{sdr}. The SDR setup includes placement in various indoor locations, such as a long indoor hallway and multiple floor levels, to evaluate reception across different environments. The SDR streams data to a laptop via USB, where AIRSPY software \cite{airspy} processes the data. We applied a Blackman-Harris 4 filters and a 2.56 Mega Sample Per Second (MSPS) quadrature sampling rate within AIRSPY. Additionally, we utilized the software’s maximum 49.6 dB RF gain, enhancing signal capture for our tests.

\subsection{Fabrication Process for a RadioGami Tag} \label{subsec:fab}
This subsection details the steps to fabricate a RadioGami tag using the Silhouette vinyl cutter and software to design, cut, and assemble the device on paper \cite{silhouette_}.

\begin{figure}[b]
\includegraphics[width=1\linewidth]{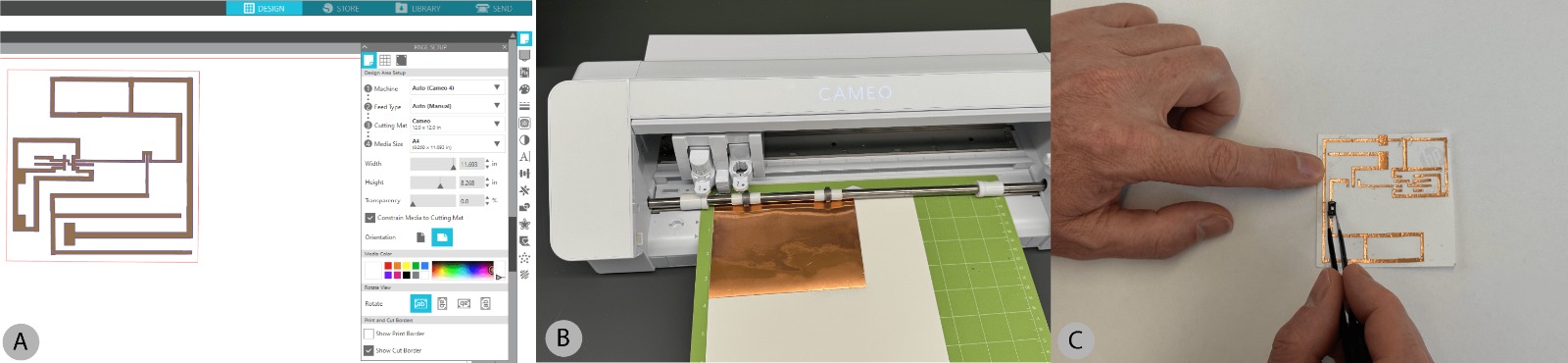}
\centering
  \caption{\textbf{Fabrication process of a RadioGami tag.} \textbf{A.} The copper trace is designed in Silhouette software. \textbf{B.} Copper tape is inserted into the Cameo 4 machine and cut. \textbf{C.} Excess copper is removed, components are added, and the final paper tag is assembled.}
\label{fig:fabprocess}
\Description[Fabrication steps of a RadioGami tag]{Detailed steps including design, cutting, component assembly, enclosure creation, and final assembly of a RadioGami tag.}
\centering
\end{figure}

The fabrication process begins with the design of the oscillator circuit. This design is created in Silhouette software, where the layout of the copper traces is defined to meet the required specifications for the circuit (Fig.~ \ref{fig:fabprocess}A). After finalizing the design, a piece of copper tape is placed onto a 64mm x 64mm card stock, which serves as a flexible substrate for the circuit. The prepared card stock with copper tape is then loaded into the Cameo 4 machine, where the machine precisely cuts the copper tape according to the design, forming the circuit traces needed for the RadioGami tag (Fig.~ \ref{fig:fabprocess}B). Once cutting is complete, excess copper is carefully removed, leaving only the desired copper traces on the card stock (Fig.~ \ref{fig:fabprocess}C). This step requires accuracy to ensure the traces remain intact and clean. With the traces in place, the next step involves attaching the circuit components, including resistors, capacitors, and a tunnel diode, to complete the oscillator circuit. These components are positioned and soldered onto the copper traces on the paper substrate, finalizing the assembly of the RadioGami tag (Fig.~ \ref{fig:circuit2}A).

The RadioGami tag consists of three active components and 24 passive components, totaling 27 components. The overall cost of the tag, including parts and fabrication, is approximately \$20.

\section{RadioGami Tag Evaluation} \label{sec:evaluation}
In this section, we evaluate the operational range of the RadioGami Tag and the impact of our novel switching circuit on the power required for its operation. We conducted the RF operational range test in a 50-meter-long indoor hallway in a laboratory building. For this setup, we placed the tag at one end of the hallway and positioned a dipole receiver antenna connected to an RTL-SDR device on the other end. Our goal was to measure the maximum transmission distance by evaluating the signal-to-noise ratio (SNR).

To determine the effective range, we measured SNR and identified the maximum distance at which the SNR dropped below 5 dB—our threshold where the signal visually blends into the noise in the receiver software (AIRSPY). The RTL-SDR first receives radio signals through its antenna, and the software (such as SDR\#) then processes these signals, converting them from the time domain to the frequency domain using a Fast Fourier Transform (FFT). The signal strength, \(P_\text{signal}\), is derived by measuring the peak power of the signal in decibels (dB), while the noise level, \(P_\text{noise}\), is calculated by analyzing the power in a nearby frequency range where no signal is present. The SNR is then calculated as:

\begin{equation}
\text{SNR (dB)} = 10 \log_{10} \left( \frac{P_\text{signal}}{P_\text{noise}} \right)
\end{equation}

\subsection{RadioGami Tag Operational Range}
To assess the RadioGami tag's long-range transmission capability with 11 and 25 photodiode configurations, we experimented to evaluate the SNR across a 50-meter hallway. In this setup, we placed the tag at one end of the hallway and moved the dipole receiver antenna connected with the SDR away from the tag in 5-meter increments. We recorded SNR values until they fell below 5 dB, defining the operational range limit for each photodiode configuration.

\begin{figure}[b]
\includegraphics[width=1\linewidth]{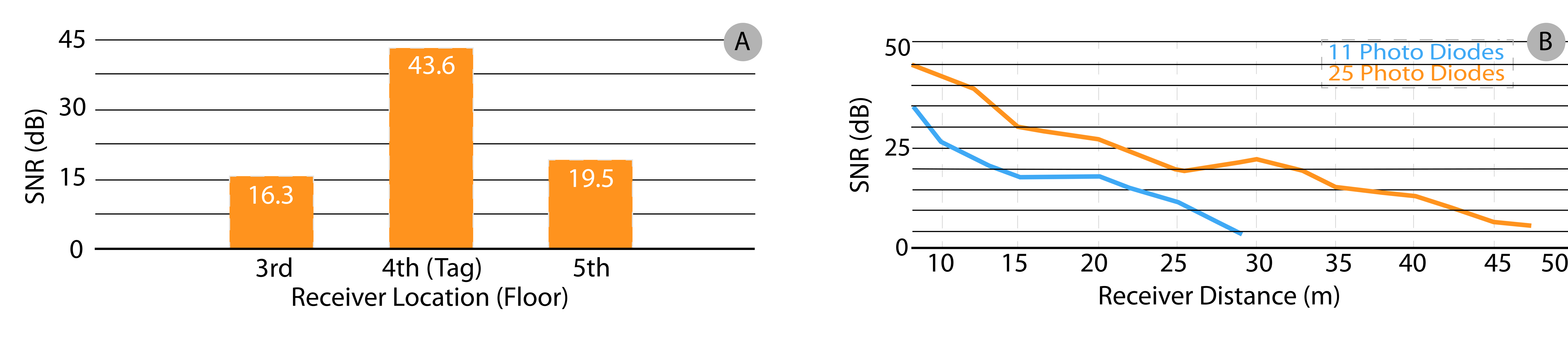}
\centering
\caption{\textbf{Building Penetration and Operational Range of RadioGami tag.} \textbf{A.} Signal-to-noise ratio (SNR) measured across three floors, showing signal strength variations as the signal penetrates different levels in a five-story building. \textbf{B.} SNR over distance for 11 and 25 photodiode configurations, illustrating the operational range and impact of photodiode count on signal strength across a 50-meter indoor hallway.}
\label{fig:rf_range}
\Description{}
\centering
\end{figure}

Fig.~\ref{fig:rf_range}B shows the results. At a near-zero distance, the 25-photodiode configuration produced a peak SNR of 45.3 dB, while the 11-photodiode configuration achieved 34.9 dB. The SNR dropped consistently with distance, reaching the minimum operational thresholds at 45.73 meters and 27.44 meters for the 25- and 11-photodiode configurations, respectively. This data demonstrates the RadioGami device's capability for long-range applications, making it suitable for use in large indoor environments, such as maker spaces, classrooms, and building scale settings.

\subsection{Multi-floor Penetration}
To evaluate the RadioGami tag's usability across obstructed indoor environments, we conducted a multi-floor experiment over five stories. In this experiment, the tag with a 25-photodiode configuration was positioned on the third floor, and the receiver was sequentially moved from the first to the fifth floor. The results, shown in Fig.~\ref{fig:rf_range}A, indicate that the third floor, where the tag and receiver were co-located, had the highest SNR at 43.6 dB. The fourth floor yielded an SNR of 19.5 dB, while the second floor recorded 16.3 dB. Although both floors are equidistant from the tag, the difference in SNR is attributed to the orientation of the tag’s antenna relative to the ground plane. The SNR values on the first and fifth floors were 4 dB and 5 dB, respectively, which were below the threshold for signal detection by the SDR. These results demonstrate that the RadioGami system can penetrate through floors despite structural obstructions.

\subsection{Impact of Intermittent Power Switching on RadioGami Tags} \label{subsec:Switch}
We evaluated the impact of intermittent power switching on our RadioGami tags, configured with 40, 25, and 11 photodiodes. This experiment aimed to understand how switching affects operational range and power consumption for each configuration under a constant light intensity of 800 lux. We measured the power drawn from the storage capacitor $C_1$ (Fig.~\ref{fig:circuit2}B) for each setup to quantify power usage. Table \ref{tab:intermittent_switch} summarizes the results, duty cycles, component values, power consumption, and maximum transmission distances.

\begin{table}[htbp]
\centering
\caption{ Comparison of RadioGami tag configurations with varying photodiode counts and \textbf{intermittent power switching} settings, showing the effects on input power consumption and maximum operational range.}
\label{tab:intermittent_switch}
\small
\begin{tabular}{cccccccccc}
\toprule
\textbf{Light} & \textbf{No.} & \textbf{Duty} & \textbf{R3} & \textbf{R4} & \textbf{CT} & \textbf{Bypass} & \textbf{Clock} & \textbf{Input Power} & \textbf{Max.} \\
\textbf{Intensity} & \textbf{of} & \textbf{Cycle} & \textbf{(M\(\Omega\))} & \textbf{(M\(\Omega\))} & \textbf{(\(\mu\)F)} & \textbf{Diode} & \textbf{(Hz)} & \textbf{(\(\mu\)W)} & \textbf{Distance} \\
\textbf{(lux)} & \textbf{PD} & \textbf{(+ve)} & & & & & & & \textbf{(m)} \\
\midrule
800 & 40 & 100\% & N/A & N/A & N/A & No & N/A & 49 & 48.78 \\
800 & 25 & 60\% & 1 & 33 & 10 & No & 60 & 35 & 45.73 \\
800 & 11 & 10\% & 1 & 33 & 10 & Yes & 24 & 16 & 27.44 \\
\bottomrule
\end{tabular}
\end{table}

\subsubsection{RadioGami Tag with 40 Photodiodes}
We used a RadioGami tag with 40 photodiodes without a switching circuit to ensure a continuous power supply. This configuration drew 49 $\mu W$ from the energy storage capacitor $C_1$ and did not require the R3, R4, or CT components due to the absence of a switching circuit. Testing demonstrated a maximum operational range of 48.78 meters, which we used as a baseline for maximum range without intermittent power switching.

\subsubsection{RadioGami Tag with 25 Photodiodes}
We configured the 25-photodiode RadioGami tag with a switching circuit to reduce power input. We incorporated a 555 timer with resistors (R3 at 1 M$\Omega$ and R4 at 33 M$\Omega$) and a timing capacitor (CT at 10 $\mu F$) to achieve a 60\% duty cycle with a 60 Hz clock frequency. This setup reduced input power consumption to 35 $\mu W$, marking a 28.6\% decrease compared to the continuous-power configuration. We measured a maximum operational range of 45.73 meters for this setup.

\subsubsection{RadioGami Tag with 11 Photodiodes}
To achieve further power reduction, we configured an 11-photodiode setup with a reduced positive duty cycle of 10\%. We added a bypass diode parallel to R4 to achieve this duty cycle with a 24 Hz clock frequency, reducing power consumption to 16 $\mu W$—a 67.3\% reduction compared to the 40-photodiode configuration. This power reduction tradeoff limited the operational range to 27.44 meters.

These results demonstrate the impact of the intermittent power switching on the input power and operational range of the RadioGami tags.

\subsection{Effect of Ambient Light Changes on Intermittent Switching}

\begin{figure}[!tb]
\includegraphics[width=1\linewidth]{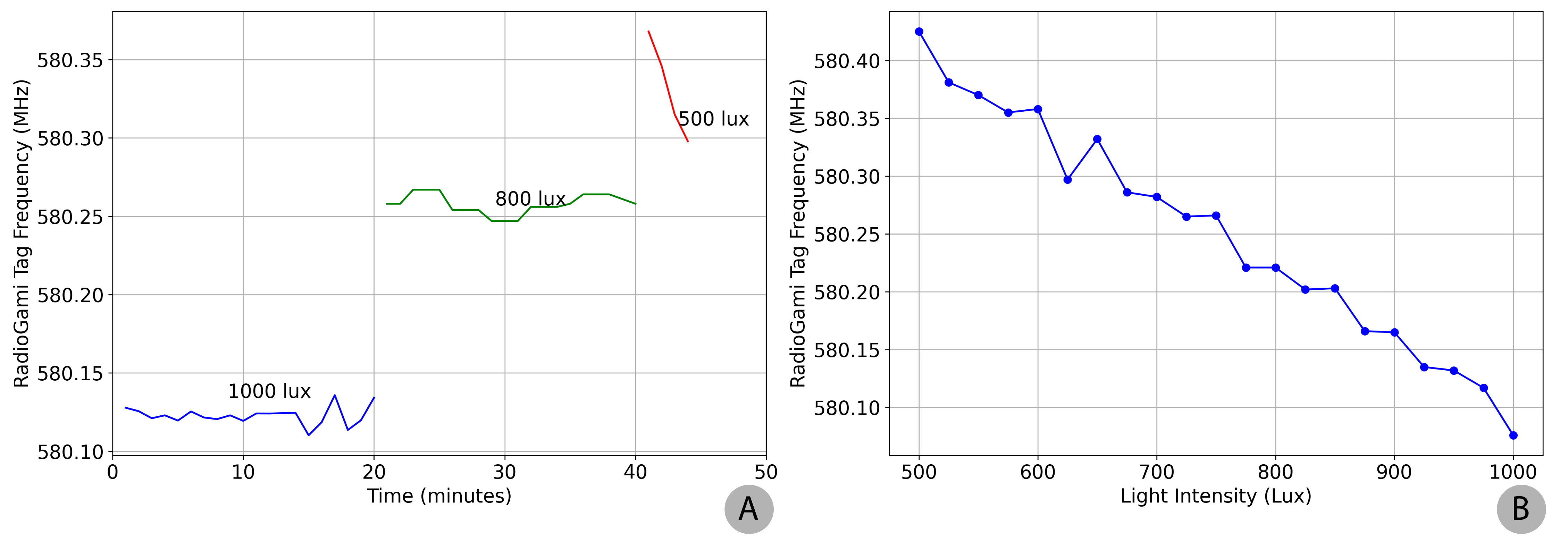}
\centering
\caption{\textbf{Effect of Light Intensity on RadioGami Tag Intermittent Switching.} 
\textbf{A.} Oscillation frequency of the RadioGami tag over
time under varying light intensities (1000 lux, 800 lux, and 500 lux). \textbf{B.} RadioGami Tag frequency as a function of light intensity from 500 to 1000 lux. The frequency decreases as light intensity increases, showing a linear relationship between light intensity and frequency response.}
\label{fig:light}
\Description{Effect of Light Intensity on RadioGami Tag Frequency. A shows receiver frequency over time with different light levels, and B shows frequency response as a function of light intensity, demonstrating the inverse relationship.}
\centering
\end{figure}

We examined how varying light intensity affects the intermittent switching behavior of the RadioGami tag. In a controlled laboratory environment, we measured the tag’s oscillation frequency under three light intensity levels: 1000~lux, 800~lux, and 500~lux (Fig.~\ref{fig:light}A). At 1000~lux, the tag maintained a stable oscillation frequency of 580.054~MHz, indicating reliable operation with sufficient harvested energy.

When we reduced the light intensity to 800~lux (20--40 minutes), the decrease in available power caused the tag’s oscillation frequency to rise. The frequency increased at a rate of 0.05~MHz per 100~lux decrease, demonstrating the tag’s sensitivity to reduced input power. At 500~lux (40--44 minutes), the harvested energy became insufficient to sustain continuous operation, producing a more substantial frequency shift. As the power deficit grew, the tag’s oscillation frequency increased rapidly until it reached a threshold where it could no longer sustain oscillations, ultimately ceasing operation due to inadequate energy.

We further analyzed the relationship between light intensity and oscillation frequency over the 500--1000~lux range (Fig.~\ref{fig:light}B). The results indicate a linear inverse correlation, with frequency increasing as light intensity decreases. A linear regression fit yielded a slope of -0.06~MHz per 100~lux, with an \( R^2 \) value of 0.98, confirming strong agreement with a linear model. These findings show that intermittent switching cannot compensate for energy deficits when the harvested power falls below the operational threshold of the tag.

These results emphasize the need to maintain higher light intensity for stable long-range operation when utilizing intermittent switching in the RadioGami tag. In environments with consistently low light levels (below 300 lux), we propose "interaction-activated" RadioGami tags, discussed in Section~\ref{sec:low_lux}, as an alternative for broader IoT applications such as environmental sensing.

\subsection{RadioGami Tag's Frequency Response to Tilt}

\begin{figure}[!tb]
    \centering
    \includegraphics[width=1\linewidth]{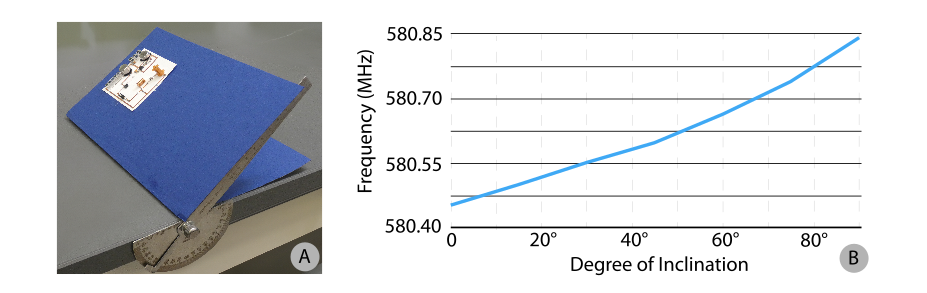}
    \caption{\textbf{RadioGami Tag Orientation Experiment.} \textbf{A.} Experimental setup where the RadioGami tag is attached to a cardstock plane and tilted in 15-degree increments relative to ambient light sources. \textbf{B.} Plot showing the frequency response of the tag at each inclination angle, illustrating the effect of tilt on oscillation frequency due to varying light intensity.}
    \label{fig:angle}
    \Description{Tag orientation experiment showing frequency changes at varying inclination angles and the apparatus used for adjusting the angle in 15-degree increments.}
\end{figure}

We experimented to measure the oscillation frequency of the RadioGami tag under varying tilt angles. In the experiment, we attached the tag to a cardstock plane and tilted it in 15-degree increments, adjusting its inclination relative to ambient light sources, as shown in Fig.~\ref{fig:angle}A. We recorded the corresponding frequency response for each tilt angle, as illustrated in Fig.~ \ref{fig:angle}B. We set the ambient light intensity to 800 lux. As we increased the tilt angle, the light intensity on the tag’s photodiodes decreased, resulting in a reduction in input voltage of the tag. This voltage reduction caused the oscillation frequency to shift upward. When the tilt angle approached 90 degrees, the incident light intensity dropped below 500 lux, lowering the voltage below the oscillation threshold and shutting down the circuit.

The frequency response increased linearly with the tilt angle, demonstrating the tag’s sensitivity to orientation relative to light. We observed an average frequency increase of 84.5 kHz per 15-degree tilt, with a standard deviation of 37.9 kHz across \( N = 30 \) trials, indicating sensitivity to changes in tilt angle.

\subsection{Deformation Study of the RadioGami Tag}
\begin{figure}[!tb]
    \centering
    \includegraphics[width=1\linewidth]{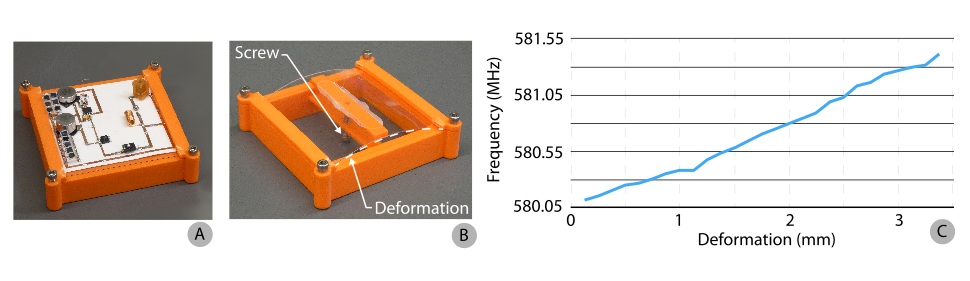}
    \caption{\textbf{RadioGami Tag Deformation Experiment.} \textbf{A.} Experimental Setup with RadioGami circuit attached.  \textbf{B.} Close-up of the device with labeled M3 screw and deformation region. \textbf{C.} Frequency change per mm of vertical deformation.}
    \label{fig:deformation}
    \Description[Deformation experiment of the RadioGami tag.]{The Fig.~ shows the deformation experiment for the RadioGami tag. A: Frequency change per mm of vertical deformation. B: Experimental setup with RF circuit attached to the tag. C: Close-up of the device with an M3 metric screw and deformation region.}
\end{figure}

\textbf{Apparatus:} The choice of paper as a circuit substrate gives the tag an advantage in using a fundamental principle of antennas: the deformation of a ground plane changes frequency. As Wang et al. \cite{wang_effect_2020} explained the linear relationship between the frequency and the amplitude of the deformation concerning the ground plane. We exploit this characteristic in mechanical paper devices by deforming the ground plane by raising/lowering a point in the ground plane with paper of different heights. Fig.~\ref{fig:deformation}A shows the final prototype we used for the deformation test.

\textbf{Experiment Procedure:} We conducted a deformation experiment using a 3D-printed apparatus, which is illustrated in Fig.~\ref{fig:deformation}. This apparatus used an M3 metric screw, which was screwed from the bottom of the apparatus to raise the crossbar height under the paper circuit. One complete turn of the M3 screw increases the height by 0.5 mm. We opted for a deformation step size of 0.125mm (¼ of a turn) to obtain more accurate data (Fig.~\ref{fig:deformation}B). During our experiment, we noticed no signal at the receiver end after 26 steps, corresponding to a height deformation of 3.25 mm to the ground plane. The tag was deformed at that moment into a maximum upside-down 'U' shape, as shown in Fig.~ \ref{fig:deformation}B. Thus, we concluded that our deformation range was limited to 26 steps of 1/4 turn. We repeated the experiment ten times and obtained comparable patterns in frequency shift, as depicted in Fig.~\ref{fig:deformation}C.

\textbf{Results:}
Fig.~\ref{fig:deformation}A reflects the change in frequency as we raised the height of the ground plane. Our experiment demonstrated a linear correlation between the deformation of the tag and the frequency change. With each ¼ turn of the M3 screw, as we increased the deformation height of the 40mm x 40mm ground plane, the oscillating frequency increased by an average of 98.74 kHz with a standard deviation (SD) of 51.3 kHz.

\section{Interactive Paper Sensors and Surfaces} \label{sec:sensors}

Researchers have long designed interactive paper sensors and surfaces. While some of these sensors are not batteryless, they typically lack wireless sensing, and current flexible batteryless wireless sensors do not support long-range sensing. Our approach is the first to embed long-range sensing capabilities into paper. In this section, we present flexible batteryless, wireless paper sensors and surfaces that utilize the RadioGami tag for micro-power, long-range interaction sensing (> 45 meters). We detail the sensing principles, fabrication techniques,  and experimental data for our paper-based rotary encoder, slider, Miura-Ori, and Kresling origami surfaces. Additionally, we demonstrate a sensor exploiting the tearing property of paper for unique interaction possibilities.

\subsection{Sensing Principle}

\begin{figure}[!tb]
\includegraphics[width=1\linewidth]{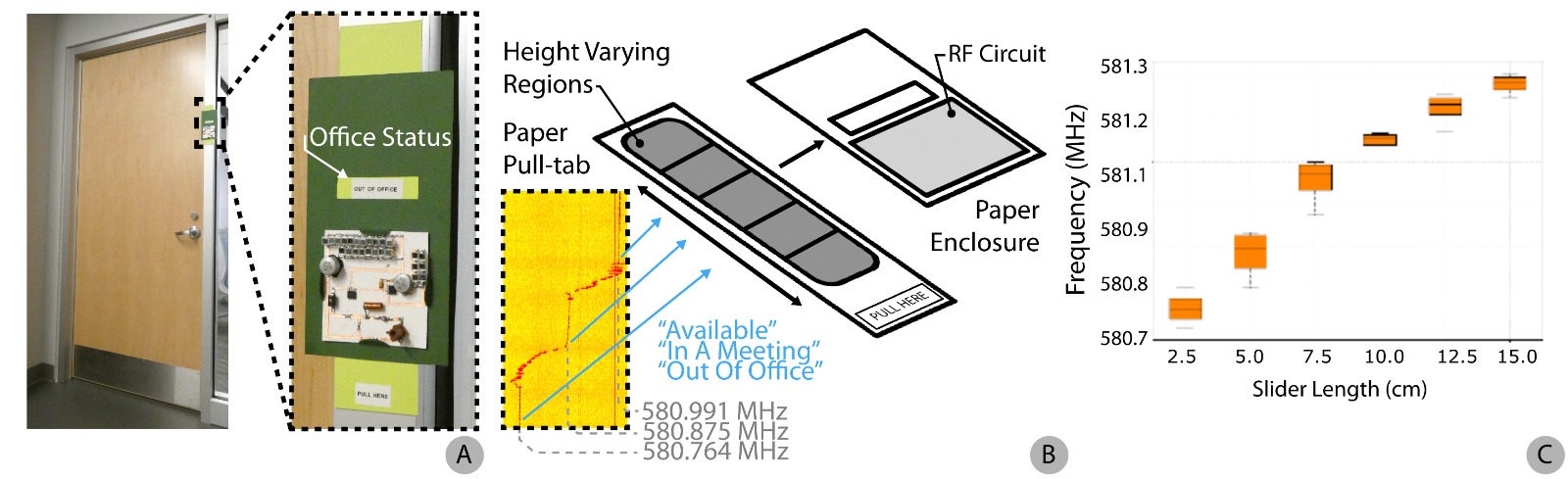}
\centering
\caption{\textbf{RadioGami Slider Application.} \textbf{A.} A sliding paper strip displays a user's office status. \textbf{B.} SDR data captures frequency changes as the slider's height varies during interaction. \textbf{C.} Frequency variation correlates with slider height and paper strip length.}
\label{fig:Slider}
\Description{RadioGami device slider application showing user interaction and resulting frequency changes along the slider.}
\end{figure}

We utilize the flexibility of our RadioGami tag to sense user activities. The paper mechanisms embedded within our sensors cause the RadioGami tag to bend and deform, leading to changes in its oscillation frequency. Initially, we attach the RadioGami tag to the slider and rotary encoder structures, as shown in Fig.~\ref{fig:Rotator}A and Fig.~\ref{fig:Slider}A. When a user interacts with these sensors, the embedded mechanisms deform the bottom surface of the RadioGami tag, which contains a ground plane for the RF transmitter. This deformation bends the ground plane, resulting in a shift in the oscillating frequency of the tunnel diode oscillator by altering its capacitance and inductance. Specifically, when the ground plane bends away from the oscillator tag, capacitance decreases, causing an increase in oscillation frequency. Additionally, changes in inductance occur due to altered magnetic field distribution, affecting impedance matching and influencing the operating point of the tunnel diode.

Next, we connect Miura-Ori (Fig.~\ref{fig:miura}A) and Kresling origami (Fig.~\ref{fig:kresling}A) sensors to the ground plane using copper tape. As users compress or expand these interactive surfaces, they affect the distributed capacitance and inductance of the RadioGami tag, causing an instantaneous shift in oscillator frequency. When the origami surface is compressed, both effective inductance and capacitance increase, shifting the oscillating frequency downward. Conversely, expanding the origami surface decreases effective inductance and capacitance, shifting the oscillating frequency upward. Our experiments in the following subsections demonstrate how these interaction techniques work and we measure frequency changes in the context of user control.

\begin{figure}[tb]
\includegraphics[width=1\linewidth]{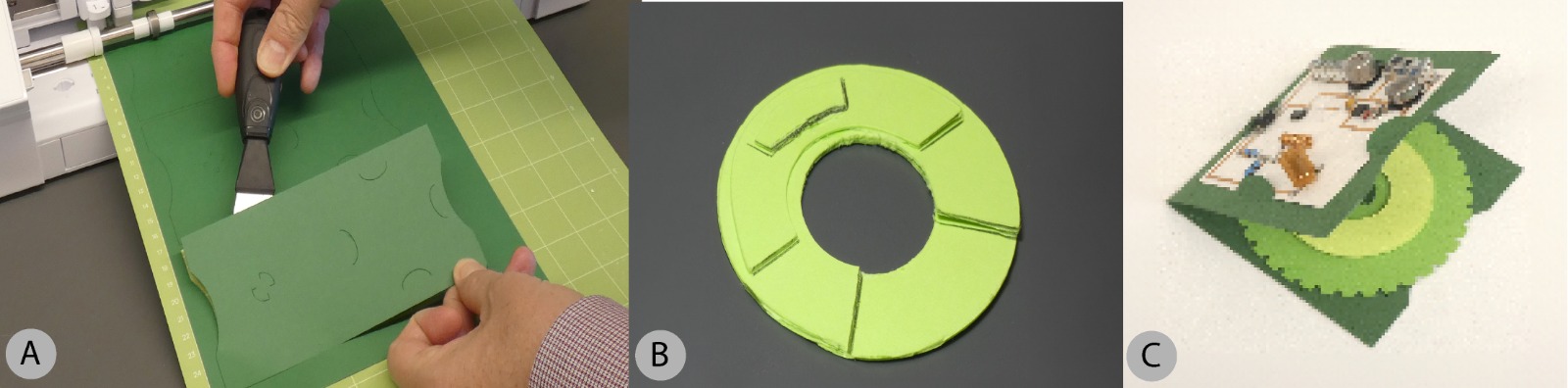}
\centering
\caption{\textbf{Rotary Encoder Assembly.} \textbf{A.} An experimenter uses a precision cutter to create the supporting structure of a Rotary Encoder. \textbf{B.} The paper wheel, designed with specific cutouts, is used inside the Rotary Encoder to facilitate smooth rotation and interaction. \textbf{C.} The RadioGami tag is attached to the Rotary Encoder sensor, ready for user interaction, allowing for RadioGami tag's oscillation frequency changes based on user input.}
\label{fig:Rotary}
\Description{Construction of the RadioGami Rotary Encoder Device with a precision cutter and attached tag.}
\end{figure}

\subsection{Rotary Encoder}
\textbf{Construction and Working Principle:}
The Rotary Encoder application, depicted in Fig.~\ref{fig:Rotary}A, features an enclosure made from folded cardstock. Inside this enclosure, a paper disc is constrained by a 360$^{\circ}$ hinge. This disc, shown in Fig.~\ref{fig:Rotary}B, includes a paper washer attached with glue, with a height variation from 0 to 2 mm. The RadioGami tag is positioned directly above this Rotary Encoder structure and secured using a paper hinge, as illustrated in Fig.~\ref{fig:Rotary}C. Notably, there is no physical or electrical connection between the tag and the sensor, allowing for easy assembly. When a user rotates the wheel, it causes the bottom surface of the RadioGami tag circuit containing the ground plane to bend. As discussed in the earlier subsection, this bending changes the distributed capacitance and inductance of the RadioGami tag, leading to a shift in its oscillation frequency.

\begin{figure}[!b]
\includegraphics[width=1\linewidth]{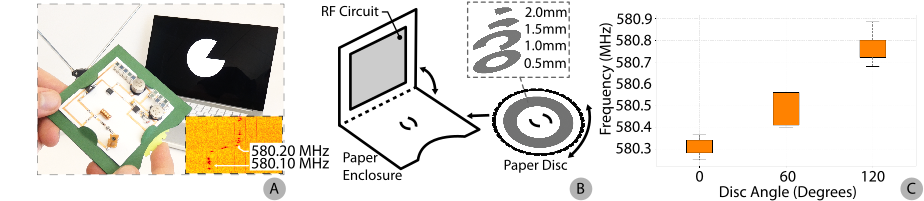}
\centering
\caption{\textbf{RadioGami Rotary Encoder Application.} \textbf{A.} User rotating a paper wheel to alter computer visualization. \textbf{B.} Spectrogram showing frequency changes during an interaction. \textbf{C.} Frequency shifts corresponding to paper wheel rotation.}
\label{fig:Rotator}
\Description{User interaction with the RadioGami Rotary Encoder Device and the corresponding frequency changes.}
\end{figure}

\textbf{Experiments and Results:}
In our experiments, we rotated the disc in increments of 60 degrees from 0 to 120 degrees and recorded tag frequencies at each stage, repeating this process 30 times to assess frequency changes over time (Fig.~\ref{fig:Rotator}A). Fig.~\ref{fig:Rotator}C summarizes our findings at different angles. At 0 degrees, the average frequency was 580.316 MHz with a standard deviation (SD) of 46.83 kHz. Rotating to 60 degrees increased the disc height by 1 mm, bending the RadioGami tag and increasing the frequency to 580.558 MHz with an SD of 112.15 kHz. At 120 degrees, a further height increase of 2 mm led to a frequency of 580.724 MHz and an SD of 81.13 kHz. We then rotated back to measure at 60 and 0 degrees again, repeating this cycle 30 times. The SD serves as a performance indicator; lower values suggest more consistent sensing via the rotary encoder.

\subsection{RadioGami Slider}
\textbf{Construction and Working Principle:}
The RadioGami slider system involves a folded cardstock enclosure secured on walls using double-sided foam tape (refer to Fig.~\ref{fig:Slider}A). Within this setup lies a sliding paper strip featuring a glued section that alters its height (Fig.~\ref{fig:Slider}B). Although not electrically linked to RadioGami tags, this mechanism influences them by bending upon movement from top to bottom over its full span (15 cm) while varying heights between 0-3 mm (Fig.~\ref{fig:Slider}B). The frequency change is continuous, but we split it into bands for consistency and repeatability measurements.

\textbf{Experiments and Results:}
Our experimental approach employed this paper slider alongside RadioGami tags to measure oscillation frequencies under varying lengths/heights for user activity detection purposes—incrementing by steps of 2.5 cm length/0.5 mm height per trial \(N=30\) as depicted in Fig.~\ref{fig:Slider}C). Once reaching 15 cm, we reversed direction and repeated measurements using identical methods.  Despite continuous changes in frequency, discrete measurements ensured repeatability for specific actions—demonstrated by standard deviation values showing precision across interactions. Initially at 2.5 cm length/0.5 mm height, frequencies averaged at 580.765 MHz (SD: 64 kHz), culminating at 15 cm/3 mm with averages at 581.279 MHz (SD: 46.2 kHz). This linear progression facilitated the development of desk status monitors that denote "Available", "In A Meeting", or "Out Of Office" statuses, as illustrated in Fig.~\ref{fig:Slider}B. We use Software-Defined Radio (SDR) data to analyze the RadioGami tag's oscillation frequencies. Specific frequency bands are mapped to each status, allowing us to determine a user's desk status based on real-time frequency readings.

\subsection{Miura-Ori Origami Interactive Surface}
\begin{figure}[!tb]
\includegraphics[width=1\linewidth]{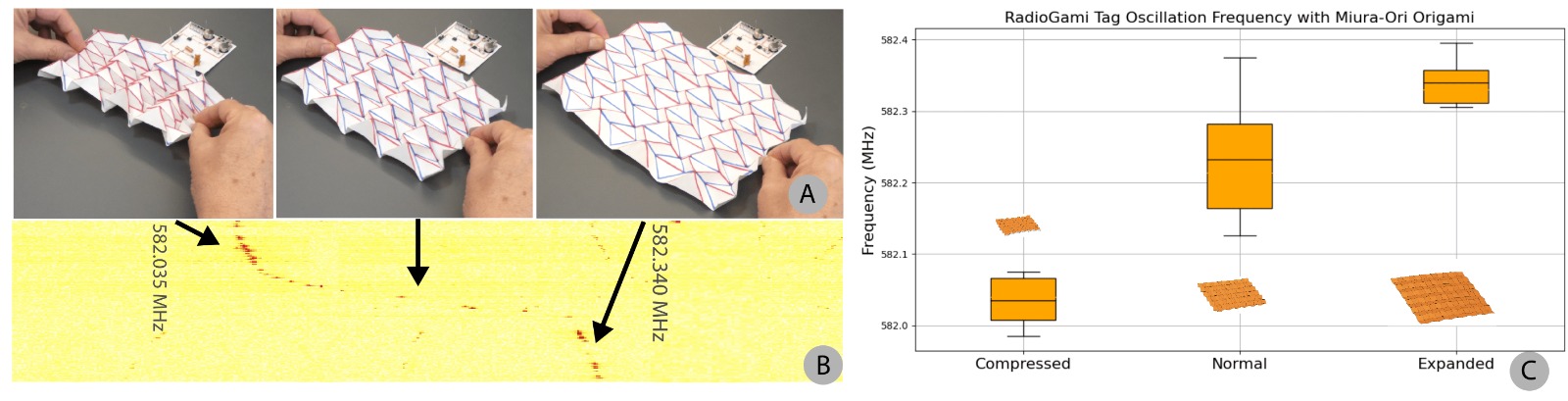}
\centering
\caption{\textbf{Interaction with a Miura-Ori sensor alters RadioGami Tag's oscillation frequency.} \textbf{A.} User compresses and expands a Miura-Ori sensor from left to right. \textbf{B.} SDR data shows frequency response changes. \textbf{C.} Box plot illustrates frequency changes across compressed, normal, and expanded states.}
\label{fig:miura}
\Description{User interaction with the Paper Radio device mountain origami and resulting frequency changes.}
\end{figure}

\textbf{Construction and Working Principle:}
The Miura-Ori origami pattern unfolds continuously through mountain and valley folds, allowing controlled expansion and contraction, as shown in Fig.~\ref{fig:miura}A. We designed an interactive frequency selective surface (FSS) using this pattern by wrapping copper tape around the peaks and valleys along the center line of the Miura-Ori surface. This copper tape connects to the RadioGami tag's ground plane. When we compress the Miura-Ori origami surface, it increases the effective inductance and capacitance of the RadioGami tag, shifting its oscillation frequency downward. Conversely, expanding the origami surface decreases these values, raising the frequency. Although these changes occur continuously (Fig.~\ref{fig:miura}B), we focused on measuring oscillation frequencies at three specific stages—normal, compressed, and expanded—to effectively quantify user interactions.

\textbf{Experiments and Results:}
We conducted experiments with the Miura-Ori interactive surface and observed distinct oscillation frequencies at each stage. In Stage 1 (normal state), we recorded the RadioGami tag's average frequency as 582.258 MHz with a standard deviation (SD) of 78.26 kHz. When we compressed it to half its length in Stage 2, the average frequency decreased to 582.059 MHz (SD = 52.71 kHz). Finally, in Stage 3 (fully expanded), it increased to 582.457 MHz (SD = 42.63 kHz). We repeated this three-stage cycle \(N=30\) times, demonstrating consistency and repeatability in dynamic sensing applications, as summarized in Fig.~\ref{fig:miura}C.

\begin{figure}[!b]
\includegraphics[width=1\linewidth]{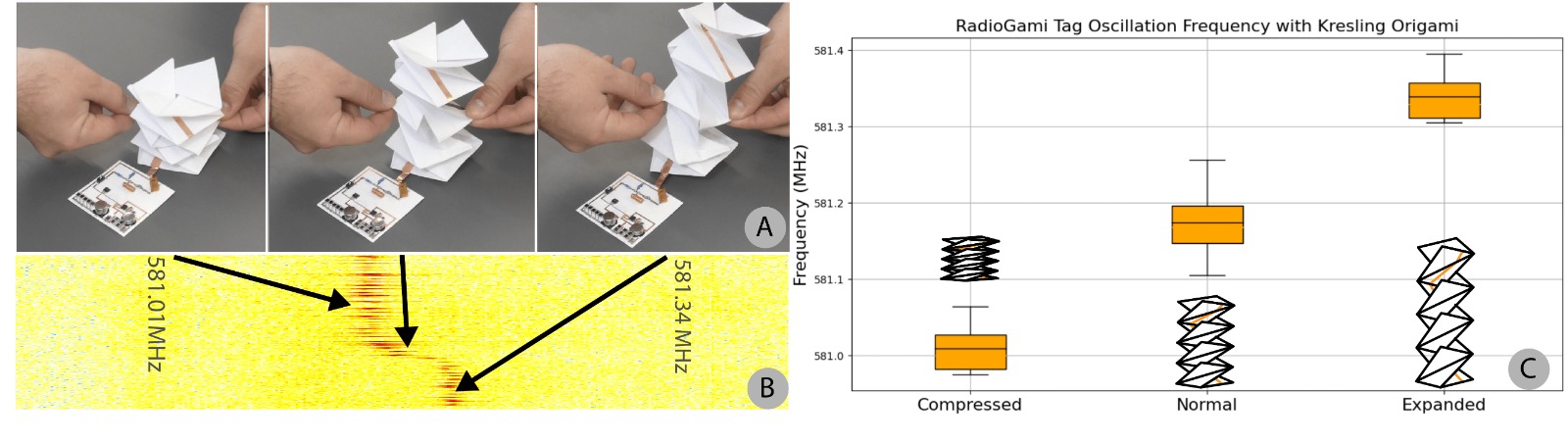}
\centering
\caption{\textbf{Interaction with a Kresling Origami sensor alters RadioGami Tag’s oscillation frequency.} \textbf{A.} Sequential images showing a user compressing and expanding a Kresling Origami sensor. \textbf{B.} SDR data illustrating frequency changes during the interaction with the origami. \textbf{C.} Box plot displaying the variation in oscillation frequency as the origami transitions between compressed, normal, and expanded states.}
\label{fig:kresling}
\Description{User interaction with the Paper Radio device spring origami and corresponding frequency changes.}
\end{figure}

\subsection{Kresling Origami Interactive Surface}
\textbf{Construction and Working Principle:}
The Kresling Origami is an interactive surface made from folded card stock, featuring a paper-based spring mechanism, as shown in Fig.~\ref{fig:kresling}A. We use a single copper tape along its height, connecting it to the RadioGami tag's ground plane. When you compress the Kresling origami surface, it increases the tag's inductance and capacitance, causing the oscillation frequency to decrease. Expanding the surface has the opposite effect, decreasing inductance and capacitance, which raises the frequency. The spring mechanism allows for compression and rebound, changing its shape and affecting the tag's circuit properties. Fig.~\ref{fig:kresling}B shows how user interactions continuously affect frequency responses.

\textbf{Experiments and Results:}
In our experiments, we assessed the RadioGami tag's oscillation frequency at three Kresling origami stages—normal, compressed, expanded—to quantify user interactions. In its normal position (Fig.~\ref{fig:kresling}C), we observed an average frequency of 581.174 MHz (SD = 58.28 kHz). Compression shifted this downward to 581.026 MHz (SD = 48.29 kHz), while expansion raised it to 581.349 MHz (SD = 60.39 kHz). We repeated this three-stage cycle 30 times, and Fig.~\ref{fig:kresling}C summarizes the experiment. These findings demonstrate the sensor's sensitivity and reliability for applications requiring tactile feedback and dynamic frequency control.

\subsection{RadioGami Package Tearing}
\begin{figure}[!htb]
    \centering
    \includegraphics[width=1\linewidth]{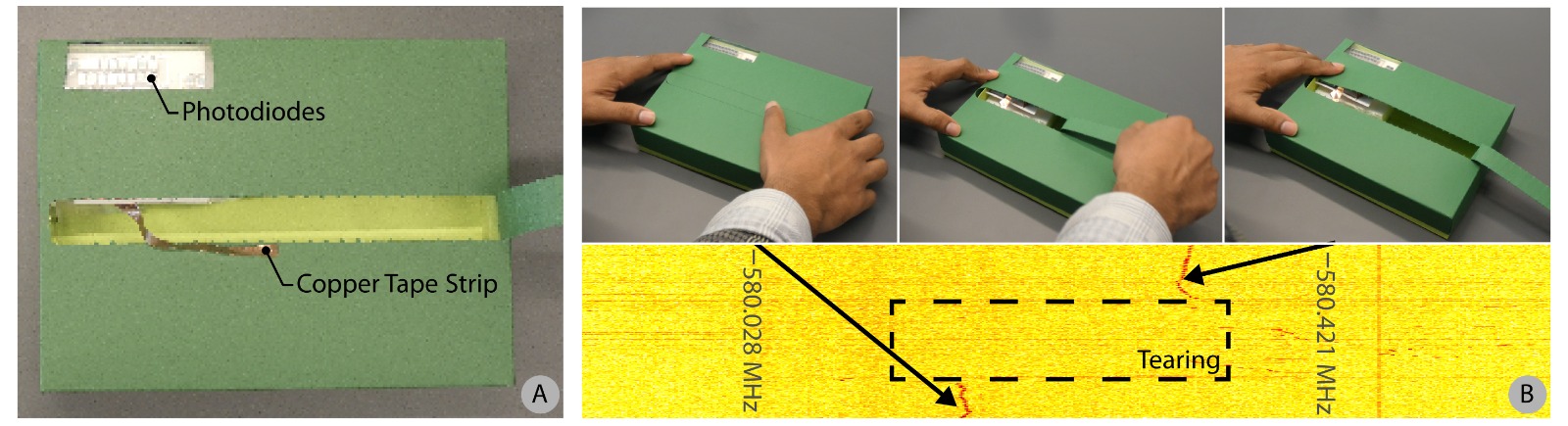}
    \caption{\textbf{RadioGami Package Tearing.} \textbf{A.} The image shows a package with an embedded RadioGami tag, including photodiodes and a copper tape strip. \textbf{B.}  The sequence illustrates the frequency response recorded as the user tears open the package, highlighting the RadioGami tag's ability to detect tampering through changes in frequency.}
    \label{fig:PackageTearing}
    \Description{Package tearing experiment with an embedded RadioGami circuit, showing the frequency response upon tearing.}
\end{figure}

We use a rectangular cardstock box with a small tear strip in the middle for package tearing detection. Inside, we embed a RadioGami tag with a hole for photodiodes to harvest ambient light, as shown in Fig.~\ref{fig:PackageTearing}A. We connect the tear strip to the RadioGami tag's ground plane using thin copper tape. When you tear the strip, the copper tape detaches from the box, reducing the circuit's inductance and capacitance and shifting the oscillation frequency upward.

Tearing part of the copper strip causes a frequency shift, depending on how much is torn. Each tear decreases the circuit's equivalent inductance and capacitance, increasing the RadioGami tag's oscillation frequency. Fig.~\ref{fig:PackageTearing}B shows a noticeable frequency shift from 580.028 MHz to 580.421 MHz upon tearing, demonstrating this method's effectiveness in detecting package tampering. These results highlight the potential for using RadioGami tags in package security, where tearing triggers a detectable frequency change, indicating tampering.

\section{Laboratory Deployment Study} \label{sec:deployment}

\subsection{Experimental Setup}
\begin{figure}[!tb]
\includegraphics[width=1\linewidth]{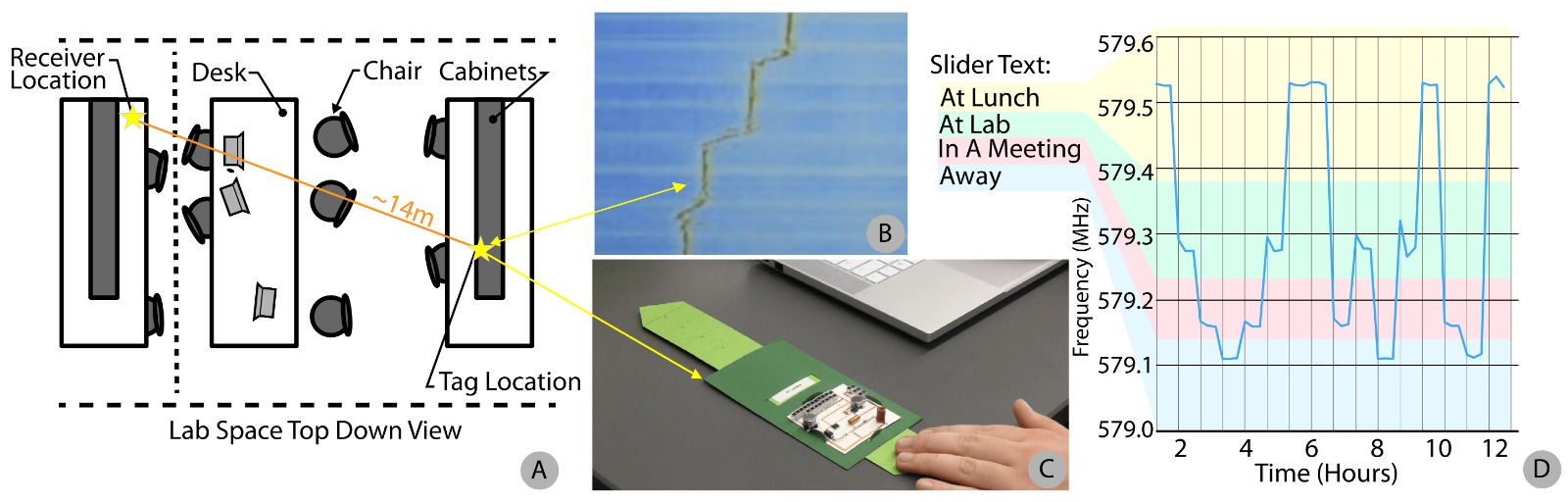}
\centering
\caption{\textbf{Deployment Study of RadioGami Tag with Slider.} \textbf{A.} Top view of the deployment study setup, showing the placement of the RadioGami tag, slider, and receiver in the lab space. \textbf{B.} Change in RadioGami tag's discretized oscillation frequency for different slider positions. \textbf{C.} Desk status monitor with printed labels for different user statuses: "At Lunch," "At Lab," "In A Meeting," and "Away," \textbf{D.} Data recorded during the deployment study shows signal variations across user interactions and positions.}
\label{fig:deployment1}
\Description{Deployment study of RadioGami Tags showing various setups and the corresponding recorded data over time.}
\end{figure}

\begin{figure}[!b]
\includegraphics[width=1\linewidth]{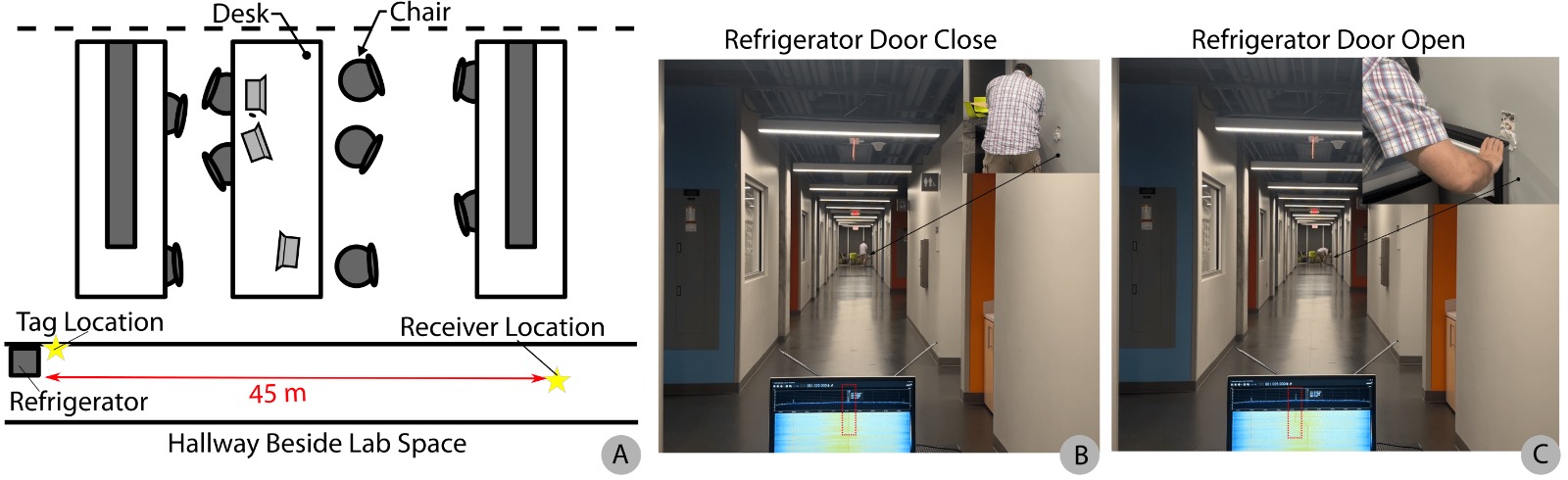}
\centering
\caption{\textbf{Deployment Study of RadioGami Tag with Kresling Origami Sensor.} \textbf{A.} Top-down view of the deployment setup showing the placement of the RadioGami tag, receiver location, and user workspaces within the lab and hallway area. The Kresling origami sensor was installed beside a refrigerator door, where its physical state changes with the door’s movement. \textbf{B.} Refrigerator door closed, illustrating the expanded state of the Kresling origami sensor. \textbf{C.} Refrigerator door open, causing the sensor to compress and affect the RadioGami tag’s oscillation frequency.}
\label{fig:deployment2}
\Description{Deployment study of RadioGami Tags showing various setups and the corresponding recorded data over time.}
\end{figure}

We conducted a laboratory scale deployment study of two RadioGami tags with paper slider and Kresling origami sensors in a controlled lab setting to evaluate their performance. We positioned one RadioGami tag on a table within the lab (Fig.~\ref{fig:deployment1}A) and placed the second tag in the hallway adjacent to the lab space (Fig.~\ref{fig:deployment2}A). We used two RTL-SDR V3 receivers with dipole antennas connected to laptops to collect data. We set up the first receiver 14 meters from the slider tag within the lab, while the second receiver, located 45 meters away in the hallway, gathered data from the Kresling origami sensor. We ensured that both tags were situated in areas with light intensities of 800 lux to maintain consistent environmental conditions Two researchers acted as users, with interactions recorded every 30 minutes between 8 am and 8 pm for four consecutive days. Before each interaction, the users informed the data logger to ensure accurate capture of all user interactions with the sensors.

\subsection{Experimental Results}
We present initial deployment study data of our RadioGami tags integrated with slider and Kresling origami sensors in Table\ref{tab:user_data}. We organized the data by showing the average oscillation frequency of the RadioGami tag, standard deviation (SD), and bandwidth (BW) of the two paper sensors.

\textbf{Slider:} We labeled the slider with four distinct positions to denote different user statuses: "At Lunch", "At Lab", "In A Meeting", and "Away". Each of these labeled positions corresponds to a unique frequency band, achieved by using a variable height strip on the pull tab. Although the RadioGami tag's frequency change is continuous, we discretized it into four distinct stages to fit this specific application scenario. We collected 96 instances of RadioGami tag frequency data (2 instances × 12 hours × 4 days) for the slider positions. The average oscillation frequency and SD remained consistent across the four labeled positions of the slider. Slider positions frequency ranged from 579.355 MHz ("At Lunch") to 579.849 MHz ("Away"). The standard deviation, SD, however, differed between the positions, reflecting the variability in user interactions. For instance, the "At Lunch" position exhibited a higher SD (67.34 kHz), suggesting variance in user interaction. In contrast, the "In A Meeting" position showed lower variability, with SD (42.13 kHz), indicating more stable interactions. Notably, all the slider positions occupy  bandwidths (BW) of 0.49 MHz with latency under 500 ms across all positions, ensuring responsiveness for real-time applications.

\begin{table}[!htb]
    \centering
    \caption{Initial deployment study of RadioGami tags integrated with the slider and Kresling sensors.}
    \small 
    \begin{tabular}{llcccccc}
        \toprule
        & Position & Average (MHz) & SD (kHz) & BW (MHz) & \\
        \midrule
        Slider & At Lunch & 579.355 & 67.34 & \\
        & At Lab & 579.594 & 49.02 & 0.49 &  \\
        & In A Meeting & 579.744 & 42.13 &  \\
        & Away & 579.849 & 58.34 & \\
        \midrule
        Kresling Origami & Expanded & 581.015 & 56.96 & 0.35 &  \\
        & Compressed & 581.365 & 48.77 & \\
        \bottomrule
    \end{tabular}
    \label{tab:user_data}
\end{table}

\textbf{Kresling origami:} We placed the Kresling origami sensor connected with a RadioGami tag beside a refrigerator door in a hallway (Fig.~\ref{fig:deployment2}). As one user opens the refrigerator door, it compresses the Kresling origami sensor shown in Fig.~\ref{fig:deployment2}B and \ref{fig:deployment2}C and affects the RadioGami tag's oscillation frequency. We recorded the oscillation frequency of the RadioGami tag when the refrigerator door is closed, opened and again closed by a user. We collected data from the Kresling origami sensor in two states: expanded and compressed. The average oscillation frequency remained consistent values of 581.365 MHz (SD: 48.77 kHz) in the compressed state. However, we observed slightly higher standard deviations for the expanded state (SD: 56.96 kHz) and average oscillation frequency of 581.015 MHz. The Kresling origami surface utilized bandwidths of 0.35 MHz and the latency was also less than 500 ms between activity actuation and RF broadcast.

\begin{table}[!htb]
\caption{Comparison of RadioGami tag with existing flexible batteryless wireless sensing tags.}
\label{tab:comparison}
\resizebox{\columnwidth}{!}{%
\begin{tabular}{lcccc}
\hline
Parameters & \textbf{RadioGami} & \textbf{PaperID} & \textbf{MARS} & \textbf{RF Bandaid} \\ \hline
Wireless Communication & RF Signal & RFID & Backscattering & Backscattering \\ \hline
Max Power & 35$\mu$W & 800$\mu$W & <1$\mu$W & 160$\mu$W \\ \hline
Max Operational Range & 45.7 m & 10 m & 15 m & 8.8 m \\ \hline
Startup Voltage & 0.25 V & N/A & 0.5V & 2.6V \\ \hline
Duty Cycle & 60\% & N/A & N/A & N/A \\ \hline
Substrate & Paper & Paper & Paper, Kapton, Polyamide & Disposable fabric band-aid \\ \hline
Receiver Model & RTL-SDR Blog V3 & EPC Gen 2 UHF RFID & Ettus Research N210 USRP & Ettus Research N210 USRP \\ \hline
Receiver Cost & <\$20 & \$200 - \$1500 & \$1838 & \$1838 \\ \hline
Surface Area of Solar Harvester& $2.84\,\mathrm{cm}^2$& N/A& $0.2\,\mathrm{cm}^2$& $45\,\mathrm{cm}^2$\\ \hline

\end{tabular}%
}
\end{table}

\subsection{Comparison of RadioGami with Other Tags}

In comparing flexible batteryless wireless sensing tags, the RadioGami tag stands out due to its impressive operational range of 45.73 meters, utilizing only 35 $\mu W$, and its low startup voltage of 0.25 V (Table \ref{tab:comparison}). These features make it ideal for micro-powered, long-range communication in power-constrained environments. While PaperID and MARS offer tags with fewer and cheaper components, they require more expensive receiver modules. Specifically, PaperID needs an RFID reader, while MARS and RF Bandaid depend on ambient carriers for backscatter communication. Despite these differences, the affordable receiver cost of <\$20 makes RadioGami an attractive option for long range applications.

\section{Additional Applications under Low Lux Conditions} \label{sec:low_lux}

\textbf{Design of Interaction-Activated RadioGami Tag:} To ensure RadioGami tags remain functional in low-lux environments (30–350 lux) and support broader IoT applications such as environmental sensing, we modified the tag to enable \textit{interaction-activated} operation. Unlike the RadioGami tags described in Section~\ref{sec:System}, which incorporate an intermittent power switching system \ref{subsec:Switching} that keeps the tag continuously "on," the interaction-activated design conserves power by keeping the tag dormant until triggered by user interaction.  

\begin{figure}[!hb]
    \centering
    \includegraphics[width=1\linewidth]{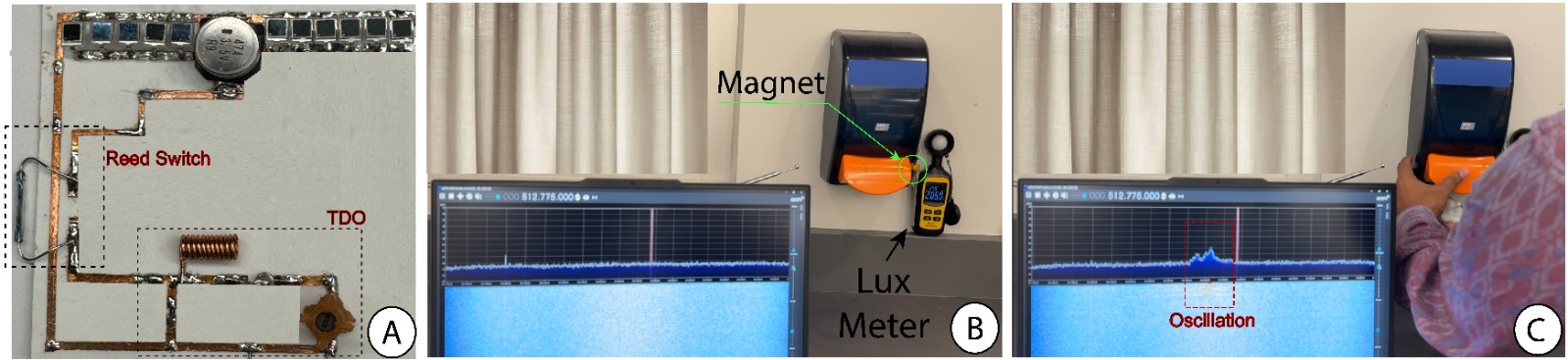}
    \caption{Activation of a RadioGami tag on a soap dispenser using a reed switch and a magnet attached to handle.  
    \textbf{A.} RadioGami tag with an integrated Reed Switch.  
    \textbf{B.} Tag attached to the side of soap dispenser with a  magnet on the push handle.  
    \textbf{C.} RF spectrogram showing activation upon dispenser press.  
    }
    \label{fig:soap}
    \Description[RadioGami tags on kitchen objects.]{Operation of RadioGami tags attached to a soap dispenser, trashcan, and toaster oven over 60 hours under variable lighting conditions.}
\end{figure}

\begin{figure}[!htb]
    \centering
    \includegraphics[width=1\linewidth]{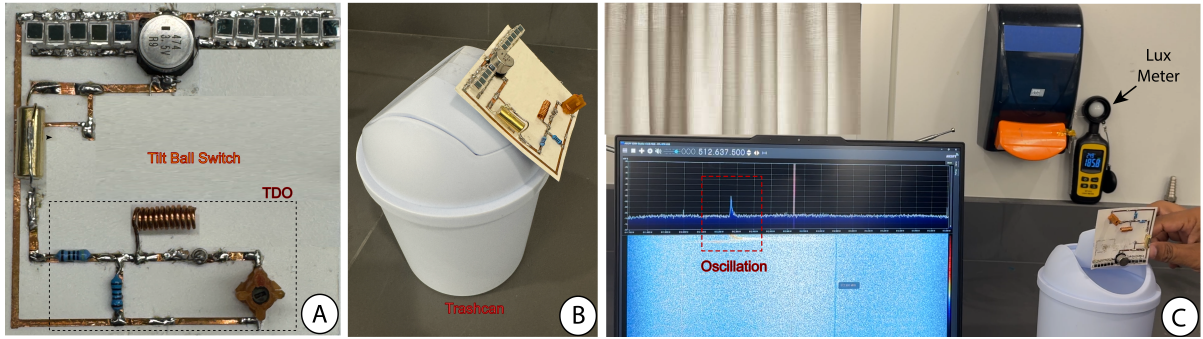}
    \caption{Activation of a RadioGami tag on a trashcan using a Tilt Ball Switch.  
    \textbf{A.} RadioGami tag with an integrated tilt ball switch.  
    \textbf{B.} Tag attached to the trashcan lid.  
    \textbf{C.} RF spectrogram showing activation when the lid tilts beyond 60 degrees, with simultaneous light intensity measurements.  
    }
    \label{fig:trashcan}
    \Description[RadioGami tags on kitchen objects.]{Operation of RadioGami tags attached to a soap dispenser, trashcan, and toaster oven over 60 hours under variable lighting conditions.}
\end{figure}

These interaction-activated RadioGami tags eliminate the need for a dedicated switching system. Instead, they incorporate passive detectors such as reed switches or tilt switches (see Fig.~\ref{fig:soap}A and Fig.~\ref{fig:trashcan}B), which momentarily power the TDO circuit by releasing stored energy from a capacitor upon interaction.  

To evaluate their effectiveness, we deployed these tags in real-world activity-sensing scenarios within a kitchen environment. Specifically, we demonstrate how RadioGami tags can monitor everyday interactions, including:  a) Soap dispenser usage, b)Trash can interactions, and c) Opening and closing of oven doors.

\textbf{Detecting Soap Dispenser Usage:}  
We attached a magnet near the soap dispenser's push button and the RadioGami tag with a reed switch (Fig.~\ref{fig:soap}A) to the side of the soap box, as shown in Fig.~\ref{fig:soap}B. When a user presses the dispenser, the magnet moves within 5 mm of the reed switch, instantly activating the tag. This interaction generates a distinct RF oscillation (Fig.~\ref{fig:soap}C), confirming activation. This setup enables passive monitoring of soap dispenser usage without requiring continuous power.

\textbf{Detecting Trashcan Interactions:}  
We integrated a tilt ball switch with the RadioGami tag to detect trashcan lid movement (Fig.~\ref{fig:trashcan}A, \ref{fig:trashcan}B). When the lid tilts beyond 60 degrees, the switch activates the tag, generating RF oscillations (Fig.~\ref{fig:trashcan}C). The tilt activation angle can be adjusted to match typical usage patterns. This setup enables passive, energy-efficient waste disposal monitoring in low-light conditions.

\begin{figure}[!htb]
    \centering
    \includegraphics[width=1\linewidth]{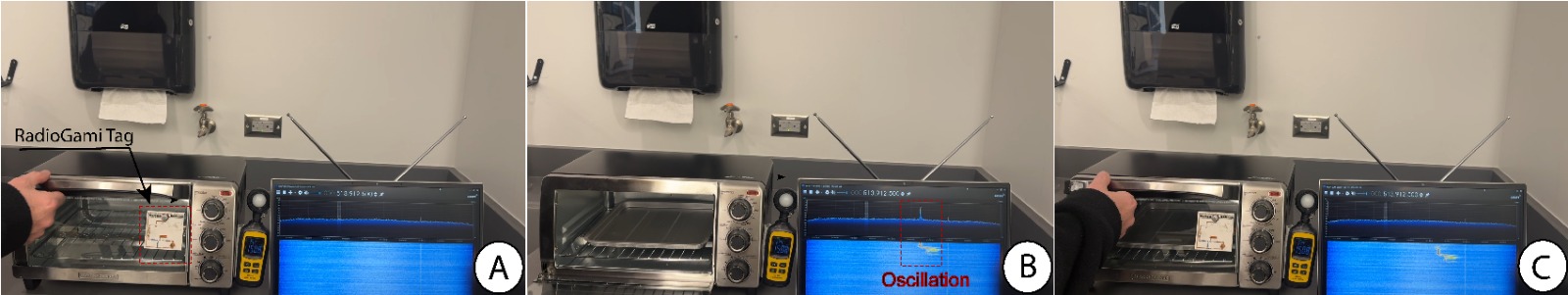}
    \caption{Activation of a RadioGami tag on a toaster oven door.  
    \textbf{A.} RadioGami tag attached to the oven door.  
    \textbf{B.} RF spectrogram showing activation upon door opening.  
    \textbf{C.} Inactive RadioGami tag after the toaster oven door being closed..  
   }
    \label{fig:oven}
    \Description[RadioGami tags on kitchen objects.]{Operation of RadioGami tags attached to a soap dispenser, trashcan, and toaster oven over 60 hours under variable lighting conditions.}
\end{figure}

\textbf{Detecting opening and closing of oven doors}  
We attached a RadioGami tag with a Tilt Ball Switch to a toaster oven door to detect open-close interactions. When the door is closed, the tag remains inactive (Fig.~\ref{fig:oven}A). Opening the door activates the tag, producing an RF oscillation (Fig.~\ref{fig:oven}B). This interaction-based activation enables passive monitoring of appliance usage. 

\subsection{Deployment Study of RadioGami Tags Under Varying Low-Lux Lighting Conditions}
\begin{figure}[tb]
    \centering
    \includegraphics[width=1\linewidth]{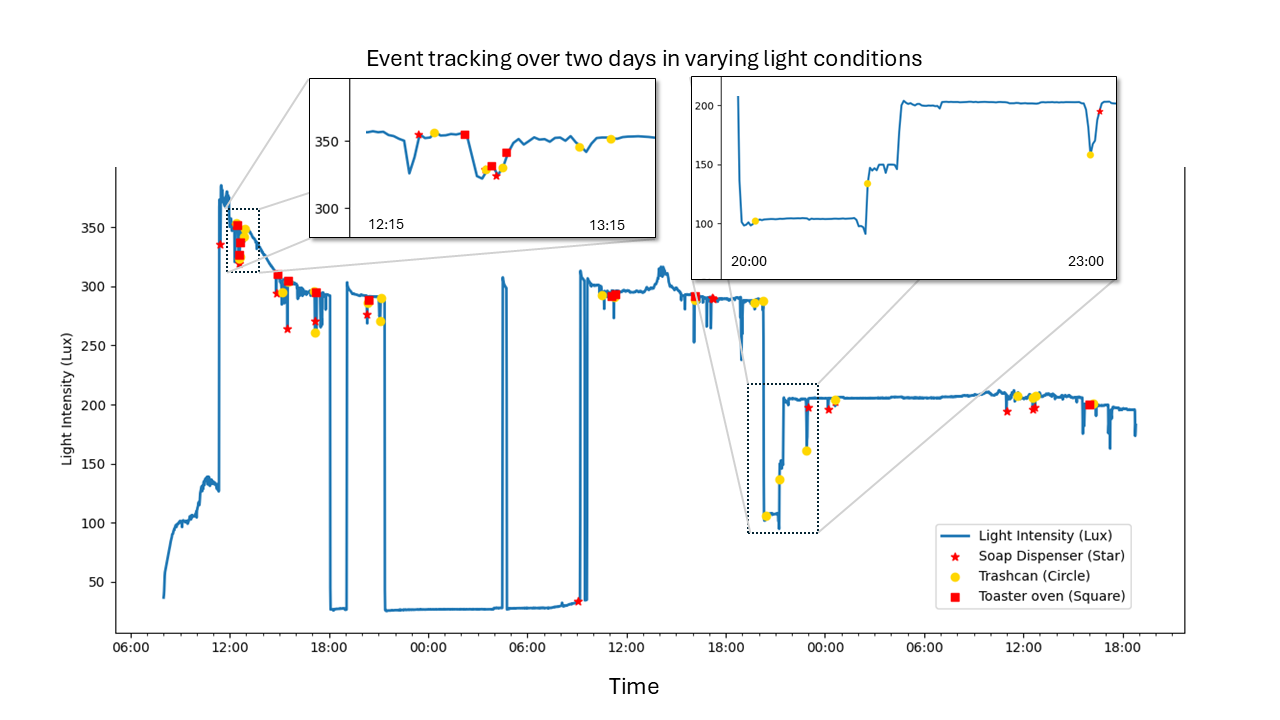}
    \caption{Event tracking of RadioGami tags over two days (60 hours) under varying light conditions. The graph illustrates light intensity (Lux) in a kitchen environment over time, with detected interactions for a soap dispenser (star), trash can (circle), and toaster oven (square). Light intensity ranges from 30 to 350 Lux, demonstrating the system's robustness in both low-light and standard-light environments. Specific events are identified based on the designated frequency band for each RadioGami tag, correlating with user interactions. The inset zooms into two specific time windows (12:15–13:15 \& 20:00-23:00) to highlight detailed event detection, showcasing the system's precision in tracking interactions despite fluctuating lighting conditions.}
    \label{fig:three_day_deployment}
    \Description[variability of RadioGami tags on different objects.]{The figure illustrates the operation of RadioGami tags attached to a soap dispenser, trash can, and oven door over 60 hours under variable lighting conditions.}
\end{figure}

We deployed the aforementioned interaction-activated RadioGami tags for 60 hours in a kitchen environment to evaluate their ability to detect user interactions and respond in varying low-light light conditions. The device testing study involved six research team members (4M, 2F) who interacted with tagged objects as part of their regular work environment. We instructed our research team members to use these tagged objects atleast once a day. The goal was to: a) Assess the robustness of RadioGami tags in naturally varying illumination conditions. b) Collect both ground-truth timestamps and system data to measure reliability. c) Confirm that RadioGami tags operated consistently across different lighting levels. The study provides strong initial evidence of system reliability under realistic, everyday conditions (as seen in Fig.~\ref{fig:three_day_deployment} above).

During the deployment, the light intensity changed between 30 and 350 lux. We used an LDR (Light Dependent Resistor) circuit and a microcontroller in proximity to the RadioGami tags to log lux levels and timestamps at 1 second interval throughout the 60 hours. We logged tag activation events whenever users interacted with the tagged objects. The system used Software Defined Radio (SDR) to record event timestamps and associated frequency response data automatically. We later processed and cross-checked with ground truths to evaluate RadioGami System performance.

\subsubsection{Results and Analysis}  
During the 60-hour deployment, we recorded 58 user-triggered events across the three tags. The system successfully detected all 19 events from the soap dispenser and all 25 from the trash can but missed 3 out of 14 events from the toaster oven, resulting in an overall failure rate of approximately 5.17\%. The missed detections were attributed to misalignment of the tilt ball switch, leading to unsuccessful switch contact. This issue stemmed from a mechanical failure of the switch itself and was not directly related to the tag. Notably, these failures occurred only with the toaster oven, while the soap dispenser and trash can maintained a 100\% detection rate.   Fig.~\ref{fig:three_day_deployment} presents an overview of the deployment study. As seen, these interactions occurred even in conditions as low as 30 lux, with a significant majority between 100 to 300 lux.

Further, we measured the tag activation time by user interaction, which varied by object type. The soap dispenser had the shortest mean activation time (0.53 s), followed by the trash can (0.68 s) and the oven door (0.9 s). The mean on-time, representing the duration the tag remained active before resetting, was highest for the oven door (6 s), followed by the trash can (3 s) and the soap dispenser (2 s). Energy consumption per interaction was 0.43\% for the soap dispenser, 1.07\% for the trash can, and 1.28\% for the toaster oven door, corresponding to differences in user interactions and activation requirements, as summarized in Table~\ref{tab:performance_metrics}.

\begin{table}[!htbp]
    \caption{{Performance Metrics of RadioGami Tags Across Different User Interactions}}
    \centering
    \small
    \setlength{\tabcolsep}{4pt}
    \begin{tabular}{lccc}
        \toprule
        Object & Mean Activation Time (s) & Mean On-Time (s) & Energy Usage (\%)/Interaction \\
        \midrule
        Trash Can    & 0.68  & 3   & 1.07  \\
        Soap Dispenser & 0.53  & 2   & 0.43  \\
        Toaster Oven Door    & 0.9   & 6   & 1.28  \\
        \bottomrule
    \end{tabular}
    \label{tab:performance_metrics}
\end{table}

\subsubsection{Operation of Multiple RadioGami Tags}  

Through this deployment study, we also validated the RadioGami system’s ability to detect simultaneous activity. Each tagged object was assigned a unique frequency band with a designated bandwidth. Using an RTL-SDR V3 receiver with a 2.4 MHz operating bandwidth, multiple tags were monitored concurrently. The tags were configured to oscillate within the 450 MHz to 600 MHz range, facilitated by an onboard variable inductor.  

\begin{figure}[!hb]
    \centering
    \includegraphics[width=1\linewidth]{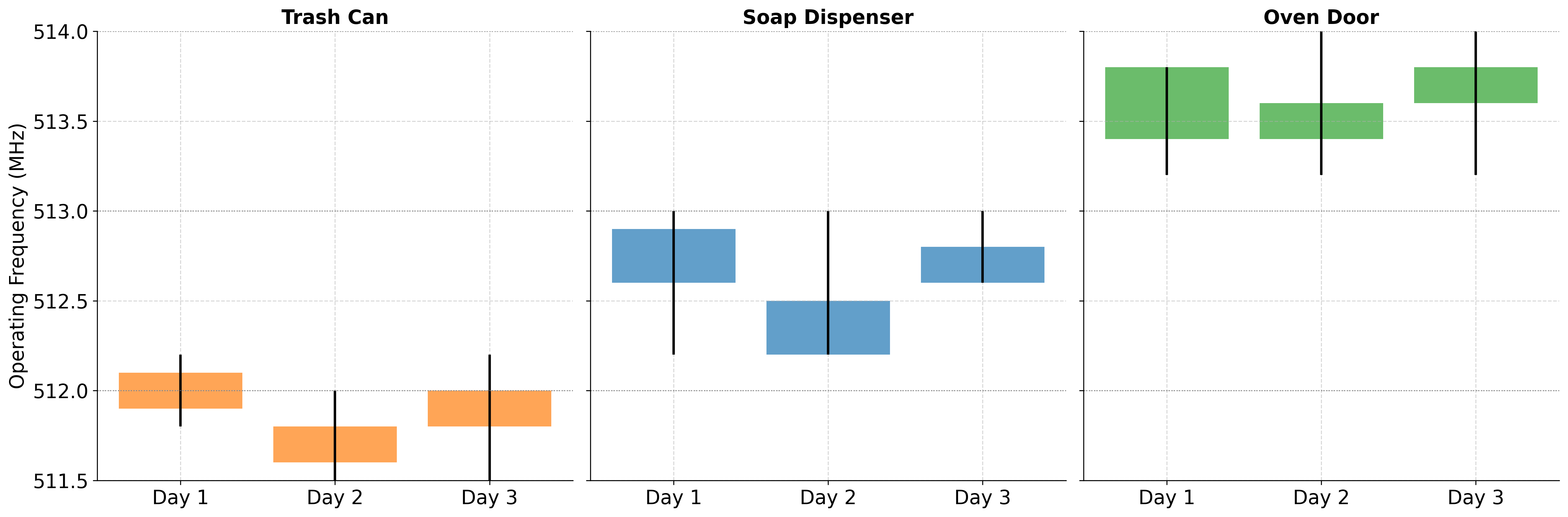}
    \caption{Frequency Variability of RadioGami Tags on Trash Can, Soap Dispenser, and Oven Door Over 60 hours. The box plots represent the distribution of operating frequencies for each tag across different days.}
    \label{fig:frequency_variability}
    \Description[Operating frequency variability of RadioGami tags on different objects.]{The figure illustrates the operating frequency variability of RadioGami tags attached to a soap dispenser, trash can, and oven door over 60 hours.}
\end{figure}

Fig.~\ref{fig:frequency_variability} illustrates the variability in operating frequencies for RadioGami tags deployed on the soap dispenser, trash can, and oven door over the 60-hour period. This analysis highlights frequency stability and overlaps across multiple tags in real-world settings.

\textbf{Frequency Stability and Variability Analysis.}
Table \ref{tab:frequency_analysis} summarizes the statistical analysis of operating frequencies for three RadioGami tags over the 60-hour deployment period. The results highlight key differences in frequency ranges, bandwidths, and stability metrics across the monitored objects.

\begin{table}[h] 
    \centering
    \caption{Statistical Analysis of Operating Frequencies for Three RadioGami Tags Over a 60-Hour Deployment.}
    \label{tab:frequency_analysis}
    \begin{tabular}{lcccccc}
        \hline
        \textbf{Object} & \textbf{Mean(MHz)} & \textbf{SD (MHz)} & \textbf{Min (MHz)} & \textbf{Max (MHz)} & \textbf{Bandwidth (MHz)} & \textbf{CV (\%)} \\
        \hline
        Trash Can & 511.71 & 0.30 & 511.0 & 512.2 & 1.0 & 0.0588 \\
        Soap Dispenser & 512.52 & 0.27 & 512.2 & 513.0 & 0.6 & 0.0517 \\
        Oven Door & 513.50 & 0.21 & 513.2 & 514.0 & 0.8 & 0.0402 \\
        \hline
    \end{tabular}
\end{table}

The Trash Can RadioGami tag operates within a frequency range of 511.0 MHz to 512.2 MHz, with a mean frequency of 511.71 MHz and a standard deviation (SD) of 0.30 MHz, occupying the bandwidth of 1.0 MHz among all tags and exhibiting the highest relative variability with a coefficient of variation (CV) of 0.0588\%. In comparison, the Soap Dispenser tag operates between 512.2 MHz and 513.0 MHz, with a mean frequency of 512.52 MHz, an SD of 0.27 MHz, and a narrower bandwidth of 0.6 MHz, resulting in a slightly lower CV of 0.0517\%. The Oven Door tag, operating at the highest frequencies (513.2 MHz to 514.0 MHz), has a mean frequency of 513.50 MHz and the smallest SD of 0.21 MHz, indicating greater stability over time; it occupies a bandwidth of 0.8 MHz and demonstrates the lowest CV at just 0.0402\%.

\textbf{Spectral Overlap and Interference Mitigation.}
Spectral analysis reveals minimal overlap between most tags' operating frequencies, reducing the likelihood of direct RF interference under normal conditions. However, some overlap between soap dispenser and trash can tags was observed in the 512–512.2 MHz range during Day 2. Correlation analysis between tag frequencies yielded a Pearson correlation coefficient of \( r = 0.12 \), indicating weak dependency among frequency trends across different tags over time. This suggests that external environmental factors do not simultaneously affect all tags due to their distinct locations and varying user interactions. We will deploy shielding materials to reduce spectral overlap and improve the reliability of frequency-based sensing systems in the future.

\section{Discussion, Limitations, and Future Work}

In this work, RadioGami demonstrated a technique for building ultra-low-power, batteryless wireless sensing devices on paper substrates for various applications. These flexible tags seamlessly integrate into daily activities, highlighting the potential of paper-based tags in advancing low-cost sensor development. The versatility of these sensors has driven rapid growth, as evidenced by recent advancements in chemical sensors, biosensors, humidity sensors, and pressure sensors. Additionally, paper-based wearable electronics represent a promising direction for future research. However, several challenges must be addressed before these wireless low-cost paper devices can achieve widespread adoption. We discuss these challenges below.

\begin{figure}[!b]
\centering
\includegraphics[width=0.70\linewidth]{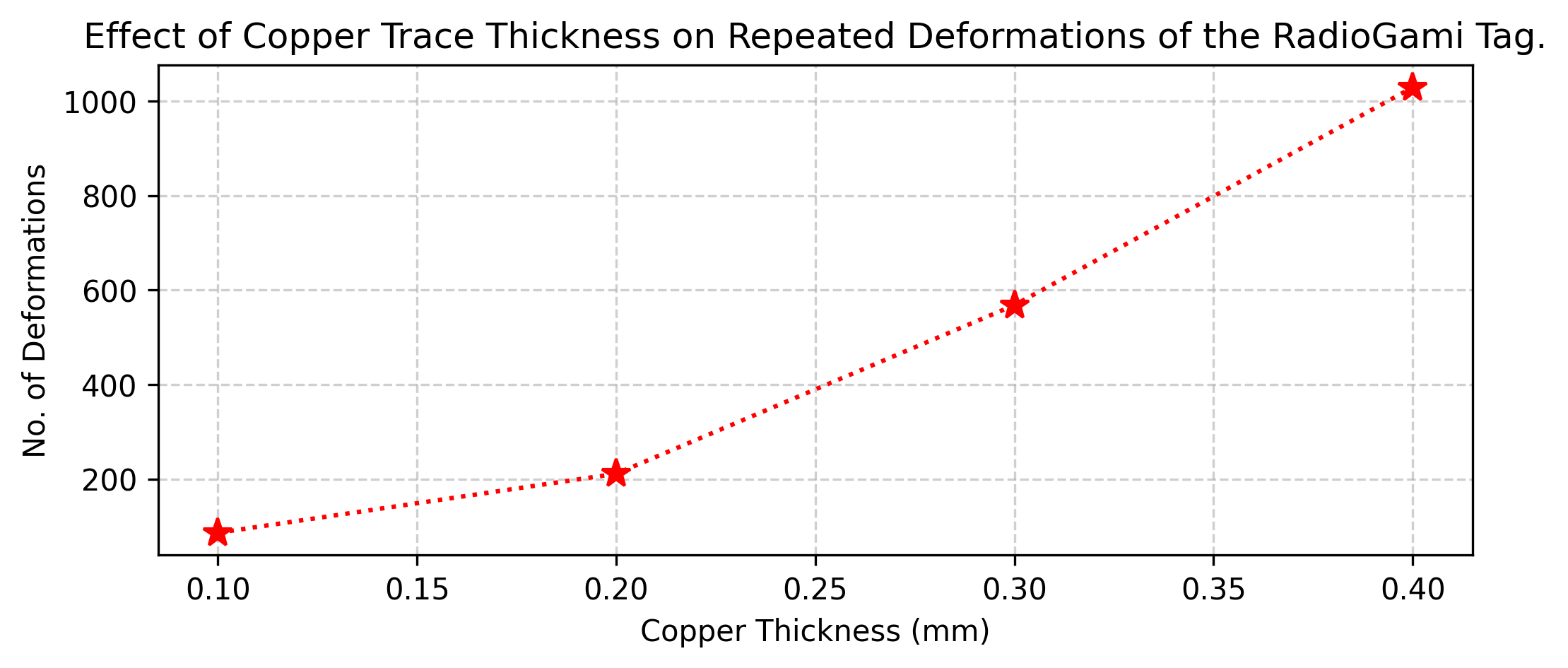}
\caption{\textbf{Effect of Copper Trace Thickness on Repeated Deformations of the RadioGami Tag.} The figure shows a positive correlation between copper thickness and durability under repeated bending. Thicker copper traces significantly increase the number of deformation cycles endured before failure, with 0.40 mm traces sustaining up to 1,000 cycles.}
\label{fig:durability}
\Description[Durability experiment of the RadioGami tag.]{The figure shows the durability experiment for the RadioGami tag under repeated deformations.}
\end{figure}

\textbf{Durability of the RadioGami Tag.} The lightweight and cost-effective paper substrate, while suitable for controlled or limited-use settings, lacks durability under repeated bending or environmental exposure, potentially compromising reliability in high-use or outdoor applications. To evaluate the durability of the RadioGami tag, we conducted a repeated bending test simulating mechanical stresses encountered during typical usage. Each cycle involved bending the tag by 10 mm and returning it to a flat state, repeated for 1,000 cycles. Visual inspections were performed every 10 cycles using a microscope to monitor degradation in the conductive pathways.

The results presented in Fig.~\ref{fig:durability} demonstrate that increasing copper trace thickness enhances durability under repeated deformations. For instance, tags with a copper thickness of 0.10 mm endure approximately 200 cycles before failure, while increasing the thickness to 0.20 mm extends this to around 400 cycles. Similarly, a thickness of 0.30 mm allows up to 700 cycles, and at 0.40 mm, the tag withstands approximately 1,000 cycles. These findings highlight the importance of optimizing copper trace thickness for applications requiring frequent bending.

\textbf{Environmental Resilience} While the RadioGami tag operates efficiently in indoor settings, its performance in outdoor environments remains a limitation. Paper-based substrates are susceptible to humidity, moisture absorption, and thermal expansion, which may degrade conductive traces and reduce sensing reliability over time. Future work should explore protective coatings, hybrid substrates, or alternative manufacturing techniques to enhance environmental resilience without compromising the tag’s flexibility.

\textbf{Fabrication and reproducibility.} Further addopting advanced fabrication techniques, such as inkjet printing with conductive ink, could improve flexibility and durability by preventing copper trace breaks during repeated use. Additionally, using surface-mount components would minimize resistive losses and improve parameter consistency, enhancing overall tag functionality.

\textbf{Wider sensing applications.} RadioGami’s long-range, batteryless wireless sensing expands paper-based sensor capabilities, making it suitable for applications like biochemical diagnostics, environmental monitoring, and wearable strain sensors. For instance, microfluidic devices on paper substrates allow controlled biochemical testing \cite{microfluidic_li_2010}, while electrochemical sensors detect heavy metals in water \cite{metals_ding2021}. Wearable paper-based strain sensors incorporate water resistance for durable applications \cite{wearable_liu_2021}, and bacteria-in-paper platforms facilitate low-cost studies of microbial dynamics \cite{hol2019bacteriapaper}. 

\textbf{Broader IoT Applications}  In this work, we demonstrated initial environmental sensing applications, such as detecting user interactions in a kitchen environment. Beyond these examples, our framework can be extended to various other applications. For instance, wearable activity sensing could be achieved by capturing wireless signal variations due to tag movement and applying signal processing and machine learning techniques.  Additionally, in medical environments, RadioGami tags can be deployed on surgical tools or equipment for passive usage tracking. Interaction-activated RadioGami tags equipped with reed or tilt switches can detect interactions such as tool pickups or repositioning. Furthermore, RadioGami-based logging can enhance inventory management by tracking disposable medical supplies, improving efficiency and reducing waste.  

While these applications extend beyond our current study, future work can focus on refining their implementations and evaluating their feasibility in real-world scenarios.

\section{Conclusion}
In this paper, we introduced RadioGami, a set of techniques for the end-to-end design of batteryless, paper-based wireless tags capable of long-range sensing (>45 meters) using ultra-low-power Tunnel Diode Oscillators (TDOs). The system detects interactions such as bending, tearing, and compression, as well as object usage, through DIY-fabricated paper sensors and origami-inspired surfaces. To remain batteryless, RadioGami devices utilize a photodiode-based energy harvesting system, while a novel intermittent power-switching circuit enables operation using scavenged energy, sustaining functionality at illumination levels as low as 500 lux. Additionally, an interaction-activated version of the RadioGami tag extends operation to low-light conditions (30–350 lux), demonstrating its suitability for energy-constrained environments.  

RadioGami’s paper-based substrate supports a wide range of interactive applications. Origami-inspired patterns such as Miura-Ori and Kresling dynamically modulate the TDO’s frequency by altering capacitance and inductance, producing frequency shifts in response to user interactions. Applications such as rotary encoders, sliders, and tamper-detection systems leverage these frequency modulations for SDR-based monitoring.  

A 60-hour deployment study confirmed the robust performance of multiple RadioGami tags in real-world environments, successfully tracking everyday interactions such as soap dispenser usage, trash can interactions, and oven door activity over substantial distances. This study provides strong initial evidence of system reliability under realistic conditions. Additionally, the recyclable substrate and reusable components enhance sustainability, supporting the development of environmentally friendly sensing technologies. This work advances ultra-low-power, low-cost sensing for IoT and ubiquitous computing, enabling the deployment of batteryless systems across diverse settings in smart environments.

\begin{acks}
The authors thank the reviewers for their valuable feedback and suggestions, which have improved the quality of this manuscript.
\end{acks}

\bibliographystyle{ACM-Reference-Format}
\bibliography{references_refined}
\end{document}